\documentclass[11pt]{article}
\usepackage[T2A]{fontenc}
\usepackage[utf8]{inputenc}
\usepackage[english]{babel}
\usepackage{graphicx}
\usepackage{wrapfig}
\usepackage{bmpsize}
\usepackage{amsmath,amsfonts,amssymb,amsthm,mathtools}
\usepackage{xcolor}
\usepackage{subcaption}
\usepackage{color}
\usepackage{indentfirst}

\definecolor{BlueGreen}{RGB}{49,152,255}
\definecolor{Violet}{RGB}{120,80,120}
\definecolor{Blue}{RGB}{0,0,255}
\definecolor{Yellow}{RGB}{0,255,51}
\definecolor{ElectricGreen}{RGB}{0, 255, 0}
\definecolor{MediumPersianBlue}{RGB}{0, 103, 165}

\usepackage[unicode, colorlinks, urlcolor=BlueGreen, linkcolor=Blue, citecolor=ElectricGreen]{hyperref}

\usepackage[left=2cm,right=2cm,top=2cm,bottom=2cm,bindingoffset=0cm]{geometry}
\numberwithin{equation}{section}

\makeatletter
\newcommand{\sgn}{ \operatorname{sgn} }

\newcommand{\ket}[1]{ | #1 \rangle }

\makeatother

\usepackage{authblk}

\title{\bf Does Schwinger's value for the current of created pairs gets modified for long and strong enough pulse?}
\author[1,2]{E.~T.~Akhmedov\thanks{\href{mailto:akhmedov@itep.ru}{akhmedov@itep.ru}}}
\author[1]{P.~S.~Zavgorodny\thanks{\href{mailto:zavgorodnij.ps@phystech.edu}{zavgorodnij.ps@phystech.edu}}}

\affil[1]{\textcolor{black}{Institutskii per, 9, Moscow Institute of Physics and Technology, 141700, Dolgoprudny, Russia}}
\affil[2]{\textcolor{black}{NRC ``Kurchatov Institute'', 123182, Moscow, Russia}}
\date{\today}

\begin{document}
\maketitle
\begin{abstract}
    We determine the parametric conditions (pulse strength and duration) under which Schwinger’s prediction for the pair-creation current --- and therefore the probability --- is significantly altered.
\end{abstract}
\tableofcontents
\newpage

\section{Introduction}\label{Introduction}

As is well known, virtual particle-antiparticle pairs are spontaneously and constantly created in a vacuum. However, an external electric field can separate these pairs by imparting sufficient energy, thereby converting the virtual particles into real ones. This process generates an electric current \cite{Schwinger:1951nm} in the presence of a constant, eternal electric field. These considerations have been extended to other types of strong background electric fields \cite{Grib:1976zw, Grib:1980aih, Zeldovich:1971mw, Nikishov, Nikishov:1969tt}—see also \cite{Gavrilov:1996pz,Kluger:1992gb,Gavrilov:2005dn,Gavrilov:2007hq,Gavrilov:2007ij,Gavrilov:2008fv,Gavrilov:2012jk,Anderson:2013zia,Anderson:2013ila,Krotov:2010ma, Akhmedov:2020dgc} for an incomplete list of related works and \cite{GribBook,Grib:1980aih} for reviews. 

Typically, in the scenarios discussed, pair creation from the vacuum is considered under conditions of a weak (in comparison with the Schwinger's critical value) background field and a low density of created particles. However, the created pairs are subsequently accelerated by the external field and emit photons, which can, in turn, decay into new pairs. This process is not forbidden in the presence of an external field because the conservation of energy does not hold strictly when backreaction on the background field is neglected. Rephrasing this, such a process happens at the cost of work performed by the external electromotive force that creates the background field.

There are, of course, also reverse processes, in which a generated pair annihilates into a photon that is subsequently absorbed by another generated particle. Such processes should play an important role when the background field is sufficiently strong and acts for a long enough duration. In our previous work, we considered such processes for a constant eternal electric field background \cite{Akhmedov:2014hfa,Akhmedov:2014doa} and extended these considerations to an electric pulse background \cite{Akhmedov:2023zfy,Akhmedov:2024rkt}.

Our eventual goal is to determine whether an avalanche of particle creation is possible for a sufficiently strong and long-lasting background electric field. More precisely, we aim to establish the conditions on the strength and duration of the background pulse required for such an avalanche to occur. At present, with achievable values of strong electric fields, this phenomenon is of purely academic interest. However, in the very early Universe --- unlike conditions created at the LHC --- there were certainly strong background gravitational fields and possibly other strong fields. A potential avalanche of particle creation in such fields could induce significant backreaction on the background gravitational field \cite{Akhmedov:2021rhq}, which may play a key role in matter and structure formation.

At this stage we address a bit less ambitious problem: we extend the considerations of \cite{Akhmedov:2023zfy,Akhmedov:2024rkt} to the case of a weak background field, using the semiclassical (WKB) approximation for the exact modes of charged fields in an electric pulse background. Our goal at this stage is to check if the Schwinger's value for the current of created pairs gets modified for long and strong enough background pulse. We calculate the tree-level current and reproduce its Schwinger's value. Then, we estimate the first and second loop corrections to it. The reason we must proceed to the second-loop correction to observe a modification of Schwinger’s current for a sufficiently long pulse is explained in \cite{Akhmedov:2023zfy,Akhmedov:2024rkt} and has a clear physical meaning: only at the second loop can we account for how radiated by pairs photons also decay into pairs, a process that can amplify the current.

For non-stationary situations involving a time-dependent background field, one should employ the Schwinger–Keldysh diagrammatic technique (for a review, see \cite{kamenev2023field}). In this approach, each field is associated with three propagators: retarded, advanced, and Keldysh. The first two characterize the spectrum of the theory, while the Keldysh propagator describes the evolution of its quantum state. As shown in \cite{Akhmedov:2014hfa, Akhmedov:2014doa} and \cite{Akhmedov:2023zfy, Akhmedov:2024rkt}, under the conditions relevant to our study, corrections to the retarded and advanced propagators are subleading compared to those for the Keldysh propagator. Thus, we are primarily interested in loop corrections to the Keldysh propagator, which describes the evolution of the quantum state and determines the resulting current.

The paper is organized as follows. In Section 2, we present the setup of the problem and define the modes for the charged scalar field in the WKB approximation. In Section 3, we calculate the tree-level current (derived from the tree-level Keldysh propagator) and correct the commonly accepted result. Additionally, in this section, we compute subleading oscillatory contributions to the tree-level current that do not grow with the pulse length. In Section 4, we explain which loop corrections to the current of created pairs dominate for a long pulse. In Section 5, we calculate the one-loop contribution to the photon’s Keldysh propagator that yields the leading correction to the current at two loops. In Section 6, we compute the leading two-loop correction to the current. In Section 7, we present our conclusions. Technical details are provided in the Appendices.

\section{The setup of the problem}

In this paper, we consider scalar electrodynamics,
\begin{equation}\label{SQED_action_initial}
S[\phi,\phi^\dagger; A^\mu]=\int d^4x\left[
|\partial_\mu\phi+ieA_\mu\phi|^2-m^2|\phi|^2
-\dfrac{1}{4}F_{\mu\nu}F^{\mu\nu}-j_{\mu}^{\text{cl}}A^\mu \right],
\end{equation}
and divide the vector potential $ A^\mu $ into the classical and quantum parts, $ A^\mu = A_{\text{cl}}^\mu + a^\mu $, where $A_{\text{cl}}^\mu$ solves the classical equations of motion with an external source $ j_{\mu}^{\text{cl}} $. The source is chosen such that the classical field is the lengthy, $eET\gg m$, but weak, $ m\gg \sqrt{eE} $, homogeneous electric pulse,
\begin{equation}\label{background_field}
A^{\text{cl}}_\mu=(0; A_1(t); 0; 0),\quad A_1(t)=ET\tanh\dfrac{t}{T}.
\end{equation}
For the quantum gauge field $ a_\mu $ we work in the Feynman gauge and perform canonical quantization in the standard way, with $\widehat{\alpha}_{\mathbf{q}\mu}^\dagger$, $\widehat{\alpha}_{\mathbf{q}'\nu}$ denoting the creation and annihilation operators, respectively. The external electric field $A^{\text{cl}}_\mu$ does not affect the free-field quantization of photons. At the same time, for the scalar field operator, we have the following decomposition into the modes:
\begin{equation}\label{phi_decomposition}
\widehat{\phi}(t,\mathbf{x})=\int\dfrac{d^3\mathbf{p}}{(2\pi)^3}\left[
\widehat{a}_{\mathbf{p}}e^{i\mathbf{px}}f_{\mathbf{p}}(t)+
\widehat{b}^{\dagger}_{\mathbf{p}}e^{-i\mathbf{px}}f^{*}_{-\mathbf{p}}(t)
\right],
\end{equation}
where the temporal part of the mode function $f_{\mathbf{p}}(t)$ solves the equation
\begin{equation}\label{modes_with_background}
\left( \partial_t^2+\Big[\mathbf{p}+e\mathbf{A}(t)\Big]^2+m^2\right)f_{\mathbf{p}}(t)=0,
\end{equation}
which follows from the classical equations of motion for the action (\ref{SQED_action_initial}).

We require the harmonic functions $f_{\mathbf{p}}(t)$ to be the in-modes, i.e. the single wave before the background electric pulse was turned on\footnote{In fact, we consider the theory to be in the standard Minkowski vacuum state before the turning on of the background electric field.}:
\begin{equation}\label{asympt_past}
f^{\text{in}}_\mathbf{p}(t/T\to-\infty)\simeq\dfrac{e^{-i\omega_{-}t}}{\sqrt{2\omega_{-}}},\quad {\rm where} \quad
\omega_\pm(\mathbf{p})=\sqrt{(\mathbf{p}+\mathbf{A}(\pm\infty))^2+m^2}.
\end{equation}
The exact in-modes of the scalar field with such an asymptotic form at past infinity are expressed through hypergeometric functions \cite{Grib:1980aih}. At the same time, the asymptotic form of the exact modes in the future infinity can be derived from the properties of these hypergeometric functions \cite{Grib:1980aih}:
\begin{equation}\label{asympt_future}
f_{\mathbf{p}}^{\text{in}}(t/T\to+\infty)\simeq\dfrac{A_+(\mathbf{p})}{\sqrt{2\omega_+}} e^{i\omega_{+}t}+\dfrac{A_-(\mathbf{p})}{\sqrt{2\omega_+}}e^{-i\omega_{+}t},
\end{equation}
where
\begin{equation}
|A_-(\mathbf{k})|^2=1+|A_+(\mathbf{k})|^2,\quad
|A_+(\mathbf{k})|^2=\dfrac{\omega_+(\mathbf{k})}{\omega_-(\mathbf{k})}\cdot
\left|\dfrac{\Gamma(\delta)\Gamma(\eta-\xi)}
{\Gamma(\eta)\Gamma(\delta-\xi)}\right|^2,
\end{equation}
and
\begin{equation}\label{xi-eta-delta}
\xi=\theta-i T \cdot \frac{\omega_{-}+\omega_{+}}{2}, \quad \eta=\theta-i T \cdot \frac{\omega_{-}-\omega_{+}}{2}, \quad \delta=1-i \omega_{-} T;
\end{equation}
\begin{equation}\label{theta-beta}
\theta=\frac{1}{2}+i \beta, \quad \beta=\sqrt{\left(e E T^2\right)^2-\frac{1}{4}}.
\end{equation}
Unfortunately, the loop integrals that appear in the expressions below cannot be calculated exactly, i.e. cannot be expressed via known special functions \cite{Akhmedov:2023zfy, Akhmedov:2024rkt}. Hence, to estimate the loop integrals it is necessary to consider an approximate form of the modes.

The approximation, which we will apply in this paper, is given by the WKB approach, in which the modes take the following form:
\begin{equation}\label{WKB-General}
f_\mathbf{k}(t)\simeq\dfrac{C_+(\mathbf{k})}{\sqrt{2\omega_\mathbf{k}(t)}}
\exp\left[i\int\limits_{t_0}^{t}\omega_\mathbf{k}(\tau)d\tau \right]+
\dfrac{C_-(\mathbf{k})}{\sqrt{2\omega_\mathbf{k}(t)}}
\exp\left[-i\int\limits_{t_0}^{t}\omega_\mathbf{k}(\tau)d\tau \right],
\end{equation}
where one can choose a convenient value for $ t_0 $. Choosing a specific value of $ t_0 $ fixes the phases of the coefficients $ C_\pm $. The form (\ref{WKB-General}) is applicable everywhere, except for small neighborhoods of the complex turning points defined by the equation
\begin{equation}\label{turning_points}
\omega^2_\mathbf{k}(t)\equiv \left[k_1+eET\tanh\dfrac{t}{T}\right]^2+\mathbf{k}_\perp^2+m^2 = 0.
\end{equation}
The complex coefficients $ C_\pm(\mathbf{k}) $ undergo jumps at $ k_1+eA_1(t)=0 $, where the WKB approximation can lose its applicability. In general, such a jump occurs during cossings of a Stokes line. We can restore the moduli of the coefficients $ C_\pm $ before and after the jump, since (\ref{WKB-General}) reproduces the asymptotics (\ref{asympt_past}) and (\ref{asympt_future}) at $ t/ T\to\pm\infty $, with
\begin{equation}\label{WKB_C_mod}
\begin{split}
|C_-(\mathbf{k},t)|^2 &= 1+|C_+(\mathbf{k},t)|^2; \\
|C_+(\mathbf{k},t)|^2 &= \dfrac{\cosh (2\pi\beta)+\cosh\left(\pi T\left(\omega_{-}-\omega_{+}\right)\right)}{2\sinh\left(\pi\omega_{-}T\right)\sinh\left(\pi \omega_{+} T\right)}\theta\Big(k_1+eA_1(t)\Big) \simeq \\
&\simeq
e^{\pi T(2eET-\omega_+-\omega_-)}\theta\Big(k_1+eA_1(t)\Big),
\end{split}
\end{equation}
where $\theta(x)$ is the theta-function, which describes the jumps at the Stockes lines. Below, to find the leading contributions to correlation functions in the limit of the lengthy pulse, we will frequently need the moduli of these coefficients.

To simplify expressions below we can Taylor expand $ \omega_\pm $ from (\ref{asympt_past}), depending on the value of the ratio of $ |k_1\pm eET|/\sqrt{\mathbf{k}^2_\perp+m^2} $. As a result, $ |C_+| $ is not exponentially suppressed only for $ |k_1|<eET-2\sqrt{\mathbf{k}^2_\perp+m^2} $, where it is given by
\begin{equation}\label{C_+-pre-gaussian}
|C_+(\mathbf{k},t)|^2\simeq\exp\left[-\dfrac{\pi(\mathbf{k}^2_\perp+m^2)}{eE}\cdot
\dfrac{1}{1-\left[\dfrac{k_1}{eET}\right]^2}\right]\theta\Big(k_1+eA_1(t)\Big).
\end{equation}
For the weak field, $ m^2\gg eE $, this expression is not negligible only in the region $ |k_1|\lesssim eET \sqrt{\frac{eE}{\pi\left(\mathbf{k}_{\perp}^2+m^2\right)}}\ll eET $, where we can perform the expansion in $ k_1/(eET) $:
\begin{equation}\label{WKB_C_refined}
|C_+(\mathbf{k},t)|^2 \simeq \exp\left[-\dfrac{\pi(\mathbf{k}^2_\perp+m^2)}{eE}\cdot\left[
1+\left[\dfrac{k_1}{eET}\right]^2\right]\right]
\theta\Big(k_1+eA_1(t)\Big)
\end{equation}
The dependence of the coefficient $ |C_+(\mathbf{k},t)|^2 $ on $ k_1 $ is necessary to know for further discussion.

In this paper, we calculate the current of created charged particles and the flux of photons produced by a lengthy electric pulse. We assume that the background field is weak, $m \gg \sqrt{eE}$, and apply the semiclassical (WKB) approximation to the exact mode functions during the pulse. The initial state of the system is taken to be the standard Minkowski vacuum for both photons and charged particles.
In the weak-field approximation, $ m\gg \sqrt{eE}$, the turning points (\ref{turning_points}) are far from the real axis in the complex $t$-plane. Consequently, the WKB approximation (\ref{WKB-General}) is valid for all real times $ t $.

\section{Refined expression for the tree-level current}

The tree-level current is given by the following expression (see e.g.\cite{Krotov:2010ma}):

\begin{equation}\label{tree-current-general}
j_1(t) = e \left\langle in \left| :\hat{\phi}^+  \overleftrightarrow{D}_1 \hat{\phi} : \right| in \right\rangle = 2 e \int \dfrac{d^3 \mathbf{p}}{(2 \pi)^3}\Big(p_1+e A_1(t)\Big)\left[\left|f_{\mathbf{p}}(t)\right|^2-\dfrac{1}{2 \omega_{\mathbf{p}}(t)}\right],
\end{equation}
where $\hat{a}_{\bf p } | in \rangle = 0$.
The tree--level expression for the Schwinger's current for $ T\lesssim t $, i.e. after the pulse, has been estimated in many places both for the scalar and standard (spinor) QED \cite{Grib:1980aih, Gavrilov:1996pz, Gavrilov:2007hq, Gavrilov:2008fv, Akhmedov:2020dgc}:
\begin{equation}\label{tree_current_renorm_finally}
j_1(t) \simeq\dfrac{E^2e^3T}{2\pi^3}\cdot\exp\left[-\dfrac{\pi m^2}{eE}\right].
\end{equation}
As we will demonstrate now, in the case of the weak field, $ m\gg\sqrt{eE} $, the pre-exponential factor in the expression (\ref{tree_current_renorm_finally}) has to be modified. In fact, we can substitute the WKB-approximation (\ref{WKB-General}) into (\ref{tree-current-general}) and obtain the following expression:
\begin{equation}
\begin{split}
j_1(t) &\simeq 2e\int\dfrac{d^3\mathbf{p}}{(2\pi)^3}\dfrac{p_1+eA_1(t)}{2\sqrt{(p_1+eA_1(t))^2+\mathbf{p}_\perp^2+m^2}}\Bigg[ |C_+(\mathbf{p})|^2+|C_-(\mathbf{p})|^2-1 \Bigg] \\
&\simeq 2e\int\dfrac{d^3\mathbf{p}}{(2\pi)^3}\dfrac{(p_1+eA_1(t))\theta(p_1+eA_1(t))}{\sqrt{(p_1+eA_1(t))^2+\mathbf{p}_\perp^2+m^2}} 
\exp\left[-\dfrac{\pi(\mathbf{p}^2_\perp+m^2)}{eE}\cdot\left[
1+\left[\dfrac{p_1}{eET}\right]^2\right]\right],
\end{split}
\end{equation}
where we have used (\ref{WKB_C_refined}). The WKB-approximation for modes allows one to estimate the tree-level current not only after the end of the pulse $ t \gtrsim T $, but also just before the end, $ t \lesssim T $. Since for $ t\sim T $,
\begin{equation}
|p_1|\lesssim \dfrac{eET}{m/\sqrt{eE}}\ll |eA_1(t)|\sim eET,
\end{equation}
we can assume that $ p_1+eA_1(t)>0 $ and $ |p_1+eA_1(t)|\gg \sqrt{\mathbf{p}_\perp^2+m^2} $. Therefore,
\begin{equation}
j_1(t)\simeq 2e \int\dfrac{d^2\mathbf{p}_\perp}{(2\pi)^3}
\int\limits_{-\infty}^{+\infty} dp_1\exp\left[-\dfrac{\pi(\mathbf{p}^2_\perp+m^2)}{eE}\cdot\left[
1+\left[\dfrac{p_1}{eET}\right]^2\right]\right].
\end{equation}

Evaluating the Gaussian integral over $ p_1 $ and the integral over $ \mathbf{p}_\perp $, which receives its main contribution from the region $ |\mathbf{p}_\perp|\sim\sqrt{eE}\ll m $, we obtain the final expression for the tree-level current in the weak-field limit:
\begin{equation}\label{tree_current_renorm_finally_weak}
\boxed{j_1(t) \simeq\dfrac{E^2e^3T}{4\pi^3\mu}\cdot\exp\left[-\pi\mu^2\right],\quad\quad \mu=\dfrac{m}{\sqrt{eE}}\gg 1.}
\end{equation}
One can see that this result is additionally suppressed by the large factor $ \mu\gg 1 $ in the denominator compared to (\ref{tree_current_renorm_finally}). This suppression was overlooked in previous estimates. However, it has a clear physical explanation: since $ \partial_t E(t)\sim 1/T $ and $ T\sqrt{eE}\gg 1 $, we can treat $ E(t) $ as approximately constant at each moment $ t $. In this case, Schwinger’s probability rate per unit four-volume is approximately
\begin{equation}
    w(t)\sim \exp\left[-\dfrac{\pi m^2}{eE(t)} \right].
\end{equation}
Since $ m\gg \sqrt{eE} $, only $ w(t) $ in a small neighborhood of $ t=0 $ (where $ E(t) $ reaches its maximum) contributes significantly to the tree-level current. The width of this neighborhood can be estimated from the following condition,
\begin{equation}
    \dfrac{\pi m^2}{eE}\cdot\dfrac{\delta E(t)}{E}\sim 1.
\end{equation}
Expanding $ E(t) $ around $ t=0 $ shows that only the region $ |t|\lesssim T/\mu $ contributes to the tree-level current. This explains why the pre-exponential factor in eq. (\ref{tree_current_renorm_finally_weak}) contains $T/\mu$ instead of $T$: the current density $j_1(t)$ along the pulse direction is proportional to $e \, w$ multiplied by the \textit{effective} pulse duration.

\subsection{Subleading oscillatory contributions to the tree-level current}

In this subsection we estimate subleading oscillatory contributions to the tree-level current that will help to understand the origin and physics of the leading loop corrections that we consider in the sections below.

In the calculation of the current (\ref{tree-current-general}) above, we have neglected the oscillatory contributions in $ |f_\mathbf{p}(t)|^2 $, which is equal to:
\begin{equation}
|f_\mathbf{p}(t)|^2 \approx \dfrac{1}{2\omega_\mathbf{p}(t)}\left[
1+2|C_+(\mathbf{p)}|^2+C_+(\mathbf{p})C_-^*(\mathbf{p})\exp\left[2i\int\limits_{t_0}^{t}\omega_\mathbf{p}(\tau)d\tau\right]
+\text{c.c.}\right],
\end{equation}
where ``c.c.'' stands for the complex conjugate expressions to those which are explicitly written. However, for $ m^2\gg eE $, we have that $ |C_+(\mathbf{p})C_-^*(\mathbf{p})|/|C_+(\mathbf{p)}|^2\sim\exp\left[\pi\mu^2/2\right]\gg 1 $. Thus, for $ \mu\ll T\sqrt{eE}\lesssim\exp\left[\mu^2 \right] $, the oscillatory contributions to the tree-level current can be comparable to (\ref{tree_current_renorm_finally_weak}) even though they are suppressed for large enough $ T $ that we consider here.

Since $ \omega_\mathbf{p}(t) $ exponentially approaches $ \omega_+(\mathbf{p}) $ as $ t/T\to+\infty $, we can use for $ t\gtrsim T $ the future infinity asymptotics (\ref{asympt_future}),
\begin{equation}\label{WKB-to-future-inf}
\left[\dfrac{C_+(\mathbf{p})C_-^*(\mathbf{p})}{2\omega_\mathbf{p}(t)}\exp\left[2i\int\limits_{t_0}^{t}\omega_\mathbf{p}(\tau)d\tau\right]
+\text{c.c.}\right]\approx
\Bigg[\dfrac{A_+(\mathbf{p})A_-^*(\mathbf{p})}{2\omega_+(\mathbf{p})}\exp\left[2i\omega_+(\mathbf{p})t\right]
+\text{c.c.}\Bigg].
\end{equation}
To perform the calculation, we need the exact expressions for $ A_\pm(\mathbf{p}) $:
\begin{equation}\label{A-pm-exact}
A_{+}(\mathbf{p}) =\sqrt{\dfrac{\omega_+(\mathbf{p})}{\omega_-(\mathbf{p})}} \cdot \dfrac{\Gamma(\delta) \Gamma(\eta-\xi)}{\Gamma(\eta) \Gamma(\delta-\xi)},
\quad
A_{-}(\mathbf{p}) = \sqrt{\dfrac{\omega_+(\mathbf{p})}{\omega_-(\mathbf{p})}} \cdot \dfrac{\Gamma(\delta) \Gamma(\xi-\eta)}{\Gamma(\xi) \Gamma(\delta-\eta)}.
\end{equation}
For $ \mu\gg 1 $, we can use the asymptotic form for the large argument for each gamma-function in (\ref{A-pm-exact}):
\begin{equation}
\begin{split}
\ln \Gamma(x+iy) &= \left[x+iy-\dfrac{1}{2}\right] \ln|y|+\sgn(y)\frac{i \pi}{2}\left[x-\frac{1}{2}\right] \\ 
&-\frac{|y|\pi}{2}-iy  
+\frac{\ln (2 \pi)}{2}+O\left[\dfrac{1}{|y|}\right]
\end{split}, \quad 1 \sim |x| \ll |y|. 
\end{equation}
After the substitution of (\ref{xi-eta-delta}) and (\ref{theta-beta}) in (\ref{A-pm-exact}), we expand $ \beta $ and $ \omega_\pm(\mathbf{p}) $ in the same way as it was done during the derivation of the the expression (\ref{C_+-pre-gaussian}). Then, since the modulus $ |C_+(\mathbf{p})| $ is exponentially suppressed for $ 1/\mu\lesssim |p_1|/(eET) $, we also expand the right-hand side of (\ref{WKB-to-future-inf}) in $ p_1/(eET) $ to quadratic order. As a result, we obtain the following expression under the integral for the subleading expression of the current:
\begin{equation}\label{tree-level-gaussian}
\begin{split}
\dfrac{2(p_1+eET)}{2\omega_+(\mathbf{p})}
A_+(\mathbf{p})A^*_-(\mathbf{p})e^{2i\omega_+(\mathbf{p})t} &\simeq
-i\cdot 2^{-2ieET^2}\left[\dfrac{\mathbf{p}^2_\perp+m^2}{(eET)^2}\right]^{-i\frac{\mathbf{p}^2_\perp+m^2}{2eE}}e^{2i\alpha eET^2} \times \\
&\times \exp\left[\dfrac{\mathbf{p}^2_\perp+m^2}{eE}\left(
-\dfrac{\pi}{2}+\dfrac{i}{2}+i\alpha \right)\right] \times \\
&\times \exp\left[ip_1 T
\left( 2\alpha+\dfrac{\mathbf{p}^2_\perp+m^2}{(eET)^2}\cdot(1-\alpha)\right)\right] \times \\
&\times \exp\left\{
\dfrac{p_1^2}{eE}\left[i+
\dfrac{\mathbf{p}^2_\perp+m^2}{(eET)^2}\left(-\dfrac{\pi}{2}
+i\alpha-\dfrac{3i}{2}-\dfrac{i}{2}\ln\dfrac{\mathbf{p}^2_\perp+m^2}{(eET)^2}\right)\right]\right\},
\end{split}
\end{equation}
where $ \alpha=t/T $. After the integration over $ p_1 $, we obtain
\begin{equation}
\begin{split}
\int\limits_{-\infty}^{+\infty} dp_1 \, (\ref{tree-level-gaussian}) &= \sqrt{\pi eE}e^{-\frac{i\pi}{4}}
2^{-2ieET^2}\left[\dfrac{\mathbf{p}^2_\perp+m^2}{(eET)^2}\right]^{-i(1+\alpha^2)\frac{\mathbf{p}^2_\perp+m^2}{2eE}}e^{i(2\alpha-\alpha^2)eET^2} \times \\
&\times \exp\left[\dfrac{\mathbf{p}^2_\perp+m^2}{eE}\left[
-\dfrac{\pi\left(1+\alpha^2\right)}{2}+\dfrac{i}{2}-i\alpha^2\left[
\dfrac{1}{2}-\alpha
\right]\right]\right].
\end{split}
\end{equation}
Finally, we integrate over $ \mathbf{p}_\perp $ and obtain the following expression for the oscillatory contribution to the tree-level electric current:
\begin{equation}
\begin{split}
 j_1^{(\text{tree,osc})}(t\gtrsim T) &\simeq 
\dfrac{e^2E}{8\pi^2}\sqrt{\pi eE}e^{-\frac{i\pi}{4}}
2^{-2ieET^2}\left[\dfrac{m}{eET}\right]^{-i(1+\alpha^2)\frac{m^2}{eE}}e^{i(2\alpha-\alpha^2)eET^2} \times \\
&\times \dfrac{\exp\left[\dfrac{m^2}{eE}\left[
-\dfrac{\pi\left(1+\alpha^2\right)}{2}+\dfrac{i}{2}-i\alpha^2\left[
\dfrac{1}{2}-\alpha
\right]\right]\right]}{
-\dfrac{\pi\left(1+\alpha^2\right)}{2}+\dfrac{i}{2}-i\alpha^2\left[
\dfrac{1}{2}-\alpha\right]-i(\alpha^2+1)\ln\dfrac{m}{eET}}+\text{c.c.} \, .
\end{split}
\end{equation}
Since $ \alpha = t/T \sim 1 $ and $ eET\gg m $, we can represent this expression as follows:
\begin{equation}\label{delta-j-tree}
\boxed{j_1^{(\text{tree,osc})}(t\gtrsim T)\simeq \dfrac{e^2E\sqrt{eE}}{4\pi\sqrt{\pi}\left[\dfrac{t^2}{T^2}+1\right]\ln\dfrac{eET}{m}}\cdot
\exp\left[-\dfrac{\pi m^2}{2eE}\left(1+\dfrac{t^2}{T^2}\right)\right]
\cos\Big[\psi_{E,m,T}(t)\Big],}
\end{equation}
where $ \psi_{E,m,T}(t) $ is some phase function, which depends on $ t $ and $ E,m,T $. One can see that, for $ t\gtrsim T $, the contribution (\ref{delta-j-tree}) is \textit{exponentially} suppressed compared to (\ref{tree_current_renorm_finally_weak}).

Let us now consider the case $ t\lesssim T $. In this situation, we cannot use the future infinity asymptotics (\ref{asympt_future}), which are applicable after the end of the pulse. Instead, one needs the exact expression for the coefficients $ C_\pm(\mathbf{p}) $ after the jump at the Stockes line:
\begin{equation}\label{C-pm-exactly}
C_\pm(\mathbf{p})=A_\pm(\mathbf{p})e^{\pm i\omega_+(\mathbf{p})t_0}
\exp\left[
\pm i\int\limits_{t_0}^{+\infty}\Big[\omega_+(\mathbf{p})-\omega_\mathbf{p}(\tau)\Big] \, d\tau    
\right].
\end{equation}
It follows straightforwardly from (\ref{C-pm-exactly}) that
\begin{equation}
C_\pm(\mathbf{p})\exp\left[\pm i\int\limits_{t_0}^{t}\omega_\mathbf{p}(\tau)d\tau\right]=
A_\pm(\mathbf{p})e^{\pm i\omega_+(\mathbf{p})t}
\exp\left[
\pm i\int\limits_{t}^{+\infty}\Big[\omega_+(\mathbf{p})-\omega_\mathbf{p}(\tau)\Big] \, d\tau    
\right].
\end{equation}
After the expansion of this expression in $ \left(\mathbf{p}^2+m^2\right)/\left(eET\right)^2 $ and $ p_1/(eET) $, we integrate over $ \tau $ to find that:
\begin{equation}
\begin{split}
\int\limits_{t}^{+\infty}\Big[\omega_+(\mathbf{p})-\omega_\mathbf{p}(\tau)\Big] \, d\tau
& = eET^2\Big[\ln(2\cosh\alpha)-\alpha\Big] +
\dfrac{\mathbf{p}^2_\perp+m^2}{2}\bigg[ \ln(2\sinh\alpha)-\alpha \bigg] + \\
& + \dfrac{\mathbf{p}^2_\perp+m^2}{2}\cdot\dfrac{p_1}{eET}\left[
1-\dfrac{1}{\tanh\alpha}\right] + \\
& + \dfrac{\mathbf{p}^2_\perp+m^2}{2}\left[\dfrac{p_1}{eET}\right]^2
\left[ \ln(2\sinh\alpha)-\alpha+\dfrac{1}{2}-\dfrac{1}{2\tanh^2\alpha} \right] + \\
&+ \mathcal{O}\left(\dfrac{\sqrt{eE}}{m}\right),
\end{split}
\end{equation}
where $ \alpha=t/T $. Then we integrate over $ p_1 $ and $ \mathbf{p}_\perp $ and obtain that, for $ t\lesssim T $, the final expression for $ j_1^{(\text{tree,osc})} $ also has the form (\ref{delta-j-tree}). 

Finally let us stress that for such $ t $ that
\begin{equation}
\dfrac{eET}{m}\ln\dfrac{eET}{m}\sim\exp\left[\dfrac{\pi m^2}{2eE}\left(1-\dfrac{t^2}{T^2}\right)\right]<\exp\left[\dfrac{\pi m^2}{2eE}\right],
\end{equation}
oscillatory contributions (\ref{delta-j-tree}) are comparable to (\ref{tree_current_renorm_finally_weak}). Below we will calculate loop corrections to the current during the pulse, $t\lesssim T$, and will see that oscillatory contributions will play the key role.

\section{General discussion of loop integrals and interrelation of the leading secular contributions to them}

We consider quantization in a spatially homogeneous state and perform the Fourier transformation along the spatial directions. We begin by examining the one-loop correction to the photon's Keldysh propagator, $G_{\mu\nu}(\mathbf{q};\tau_1,\tau_2)$, focusing on times $ |\tau_{1,2}|\lesssim T $, i.e. on the behavior of the correlation functions during the pulse.

The reason for that is as follows. First, before the pulse is turned on, the electric current, level populations and anomalous quantum averages are identical to those in empty state. In fact, we take the initial state of the theory to be the in-ground Fock state $ \ket{\text{in}} $, which coincides with the Poincare invariant state, i.e. the standard Minkowski vacuum.

Second, after the background field is switched off, the theory reduces to ordinary scalar QED (without external fields), but with a nontrivial initial state. As we will see, non-trivial level populations and anomalous averages for both photons and charged particles can be generated during the pulse. The subsequent evolution of such a state in QED without background fields constitutes a distinct physical scenario, with its own specific features and academic interest --— related to thermalization processes \cite{Akhmedov:2021vfs}, but not to particle creation by the background electric field. Therefore, in the present work, we focus on the kinetics \textit{during} the electric pulse, which is why all external time arguments of the correlation functions considered below are taken to be less than $T$.

\begin{figure}[h]
\center{\includegraphics[width=0.6\linewidth]{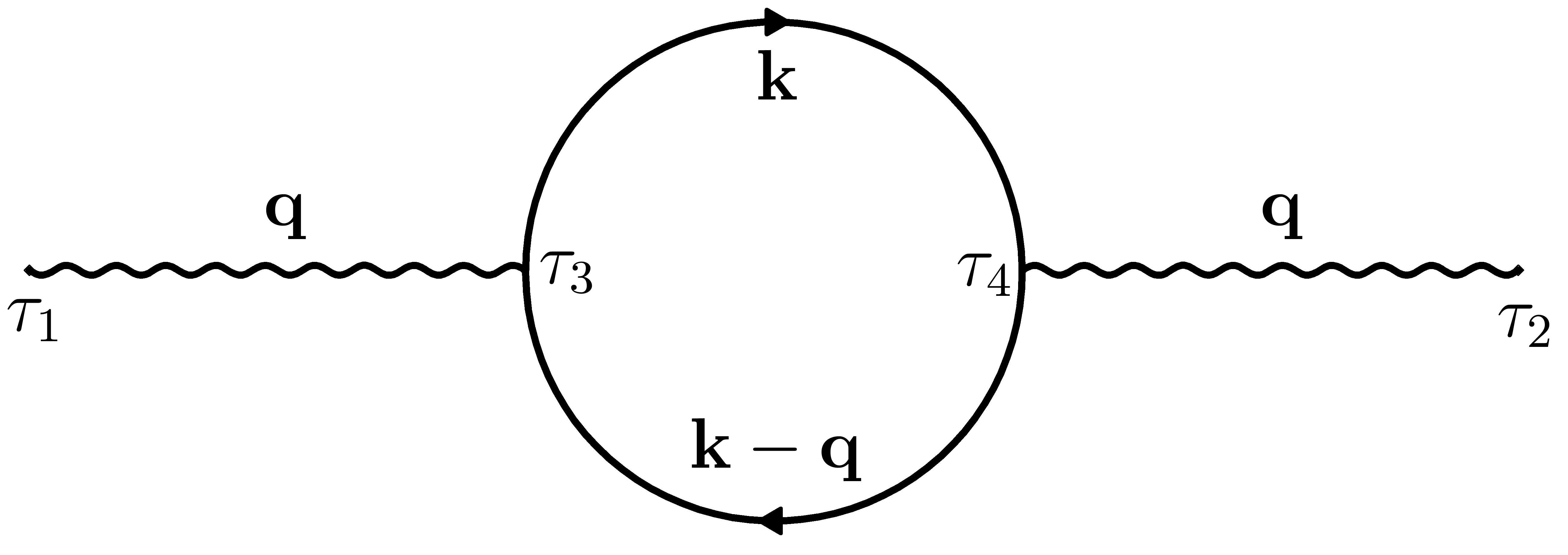}}
\caption{One-loop correction to the photon's propagator. In the Schwinger-Keldysh technique one has a sum of diagrams of such a form with vertexes attributed to different sides of the Keldysh's ``closed'' time contour.}
\label{image_dia_photon_0}
\end{figure}

In \cite{Akhmedov:2023zfy}, we estimated the one-loop correction to the photon's Keldysh propagator, $G_{\mu\nu}(\mathbf{q};\tau_1,\tau_2)$, shown in Fig. \ref{image_dia_photon_0}. In the limit\footnote{We consider such a limit to single out the destiny of the state of the theory by the end of the pulse, which will become clear from the discussion below.}. 

\begin{equation}\label{Thelimit} 
|\tau_1-\tau_2|\ll|\mathcal{T}| \lesssim T,\,\,\, {\rm where} \,\,\, \mathcal{T}=(\tau_1+\tau_2)/2, 
\end{equation}
the correction takes the form
\begin{equation}\label{K_photon}
\Delta G^K_{\mu\nu}(\mathbf{q};\tau_1,\tau_2)\simeq n^+_{\mu\nu}(\mathbf{q},\mathcal{T})
\dfrac{e^{i|\mathbf{q}|(\tau_1-\tau_2)}}{2|\mathbf{q}|}+
 n^-_{\nu\mu}(\mathbf{q},\mathcal{T})
\dfrac{e^{-i|\mathbf{q}|(\tau_1-\tau_2)}}{2|\mathbf{q}|},
\end{equation}
where
\begin{equation}\label{n=munu-creation-annihilation}
(2\pi)^3 n^-_{\mu\nu}(\mathbf{q}) \, \delta \left(\mathbf{q} - \mathbf{q}'\right) = \left\langle \widehat{\alpha}_{\mathbf{q}\mu}^\dagger
\widehat{\alpha}_{\mathbf{q}'\nu}\right\rangle,\quad
(2\pi)^3 n^+_{\mu\nu}(\mathbf{q}) \, \delta \left(\mathbf{q} - \mathbf{q}'\right) = \left\langle \widehat{\alpha}_{-\mathbf{q}\mu}^\dagger
\widehat{\alpha}_{-\mathbf{q}'\nu}\right\rangle,
\end{equation}
and $\widehat{\alpha}_{\mathbf{q}\mu}^\dagger$, $\widehat{\alpha}_{\mathbf{q}'\nu}$ are photon's creation and annihilation operators, respectively. The quantum expectation values in these expressions are taken with respect to the evolved initial state. Due to the presence of the background field these expectation values are non-zero, in contrast to the Poincare invariant state and evolution without strong external fields.

Meanwhile, the tree-level photon's Keldysh propagator is
\begin{equation}\label{K_photon1}
G^K_{\mu\nu}(\mathbf{q};\tau_1,\tau_2) = \dfrac{\cos\Big[|\mathbf{q}|(\tau_1-\tau_2)\Big]}{2|\mathbf{q}|},
\end{equation}
arising purely form zero-point vacuum fluctuations. Hence, $ n^\pm_{\mu\nu}(\mathbf{q})$ are photon level populations, which, being zero at the initial Cauchy surface, are generated by the background electric pulse, as we will see in a moment. Thus, photons are generated on top of the apparent creation of charged particles \cite{Akhmedov:2014doa}, \cite{Akhmedov:2014hfa}, \cite{Akhmedov:2023zfy}.

For $ n^+_{\mu\nu}(\mathbf{q},\mathcal{T}) $ we obtain the following integral expression \cite{Akhmedov:2023zfy}:
\begin{equation}\label{n_munu_initial}
\begin{aligned}
n^+_{\mu \nu}(\mathbf{q}, \mathcal{T}) & \simeq e^2 \int\limits_{-T}^{\mathcal{T}} d \tau_3 \int\limits_{-T}^{\mathcal{T}} d \tau_4 \frac{e^{-i|\mathbf{q}|\left(\tau_3-\tau_4\right)}}{2|\mathbf{q}|}\times \\ 
&\times\int \frac{d^3 \mathbf{k}}{(2 \pi)^3} f_{\mathbf{k}}\left(\tau_3\right) \overleftrightarrow{D}_\mu f_{\mathbf{k}-\mathbf{q}}\left(\tau_3\right) \cdot f_{\mathbf{k}}^*\left(\tau_4\right) \overleftrightarrow{D}_\nu^{\dagger} f_{\mathbf{k}-\mathbf{q}}^*\left(\tau_4\right),
\end{aligned}
\end{equation}
where
\begin{equation}
f_{\mathbf{k}}\left(\tau_3\right) \overleftrightarrow{D}_\mu f_{\mathbf{k}-\mathbf{q}}\left(\tau_3\right)= D_\mu f_{\mathbf{k}}\left(\tau_3\right)\cdot f_{\mathbf{k}-\mathbf{q}}\left(\tau_3\right)-f_{\mathbf{k}}\left(\tau_3\right) D_\mu^{\dagger}f_{\mathbf{k}-\mathbf{q}}\left(\tau_3\right),
\end{equation}
and similarly for $\overleftrightarrow{D}_\nu^{\dagger}$. The expression for $ n^-_{\mu\nu}(\mathbf{q},\mathcal{T}) $ can be obtained from (\ref{n_munu_initial}) by replacing $ \overleftrightarrow{D}_\mu\mapsto\overleftrightarrow{D}_\mu^\dagger $ and $ \overleftrightarrow{D}_\nu^\dagger\mapsto \overleftrightarrow{D}_\nu $. Equivalently, we can perform the change $ \mathbf{k}'=\mathbf{k}-\mathbf{q} $ in the integral over $ \mathbf{k} $ in (\ref{n_munu_initial}) to obtain that
\begin{equation}
n^-_{\mu\nu}(\mathbf{q},\mathcal{T})=n^+_{\mu\nu}(-\mathbf{q},\mathcal{T}),
\end{equation}
which agrees with (\ref{n=munu-creation-annihilation}). Since in the presence of the external electric field photons can be emitted by created charged particles, which, in turn, are accelerated along the direction of the field, one can expect that $ n^+_{\mu\nu}(\mathbf{q},\mathcal{T}) $ actually depends not only on the absolute value of $ \mathbf{q} $, but also on its direction. In our previous works \cite{Akhmedov:2023zfy}, we did not emphasize this point, as we were primarily concerned with whether the expression (\ref{n_munu_initial}) exhibits secular growth in $ T $, rather than with the explicit density of the created photons. In this note, however, since we aim to directly calculate the contribution of these expressions to the second-loop correction to the current, we have to account for such dependence.

Another useful symmetry is $ n^\pm_{\nu\mu}(\mathbf{q},\mathcal{T})=n^{\pm *}_{\mu\nu}(\mathbf{q},\mathcal{T}) $, from which it follows that
\begin{equation}
\left[\Delta G^{K}_{\mu\nu}(\mathbf{q};\tau_1,\tau_2)\right]^*=\Delta G^{K}_{\mu\nu}(-\mathbf{q};\tau_1,\tau_2)=\Delta G^{K}_{\nu\mu}(\mathbf{q};\tau_2,\tau_1).
\end{equation}
In \cite{Akhmedov:2023zfy} we showed that (\ref{n_munu_initial}) grows linearly with $ T $ in the limit (\ref{Thelimit}), while the anomalous quantum expectation value\footnote{The anomalous quantum expectation value is defined as $(2\pi)^3 \kappa_{\mu\nu}(\mathbf{q}) \, \delta \left(\mathbf{q} + \mathbf{q}'\right) = \left\langle \widehat{\alpha}_{\mathbf{q}\mu}
\widehat{\alpha}_{\mathbf{q}'\nu}\right\rangle$.} $ \kappa_{\mu\nu}(\mathbf{q},\mathcal{T}) $ does not grow with $ T $. In fact, anomalous quantum expectation value is also generated, but the generated value is suppressed by fine-structure constant $e^2$ and is not enhanced by the large factor $T$. For this reason, we neglect the anomalous expectation values in (\ref{K_photon}). Physically, this means that the initial vacuum state for photons that we consider here remains to be the vacuum in the future.

Then, in \cite{Akhmedov:2024rkt} it was demonstrated that the secular growth of $ n^\pm_{\mu\nu}(\mathbf{q})$ leads to the growth with $T$ of the second-loop correction to the current of created pairs of charged particles. The physical explanation is straightforward: as mentioned above, created charged particles accelerate and radiate photons. These radiated photons can then decay into pairs in the background electric field, which in turn enhances the current of the created charged pairs. Essentially, the integrals considered in this paper are components of collision integrals for particle kinetics in a background field. They are nontrivial and differ significantly from the standard Boltzmann-type collision integrals because we consider exact modes in the background fields rather than plane waves.

To find the second-loop correction to the current, we have to estimate the above one-loop correction to the photon's Keldysh propagator using the approximate WKB form of the modes of the charged field. Specifically, we single out such contributions to (\ref{n_munu_initial}), which provide the leading corrections to the current in the second-loop correction. The latter are those which have the highest power of growth with $T$ and do not oscillate with $ \mathbf{q} $. Below we will demonstrate that only the photons in the limit of soft momenta, $ |\mathbf{q}|=\mathcal{O}(1/T) $, but with \textit{all possible} values of the time arguments of the photon's Keldysh propagator, $ \tau_{1,2} $, (not restricted to the region $ |\tau_1-\tau_2|\ll |\tau_1+\tau_2| $), give the leading contribution to the second-loop correction to the current. For this reason, in the present paper, we calculate one-loop integral only for such soft photons. Since we go beyond the limit (\ref{Thelimit}), $ n_{\mu\nu} $ and $ \kappa_{\mu\nu} $ individually lose their direct physical interpretation. Therefore, we should consider their combined contribution to the entire correction to the Keldysh propagator, $ \Delta G^{K}_{\mu\nu} $. This is the subject of the next section.
 
In our previous paper \cite{Akhmedov:2024rkt}, we have obtained the integral expression for the two-loop correction to the electric current, shown in Fig. \ref{image_dia_photon}:
\begin{figure}[h]
\center{\includegraphics[width=0.6\linewidth]{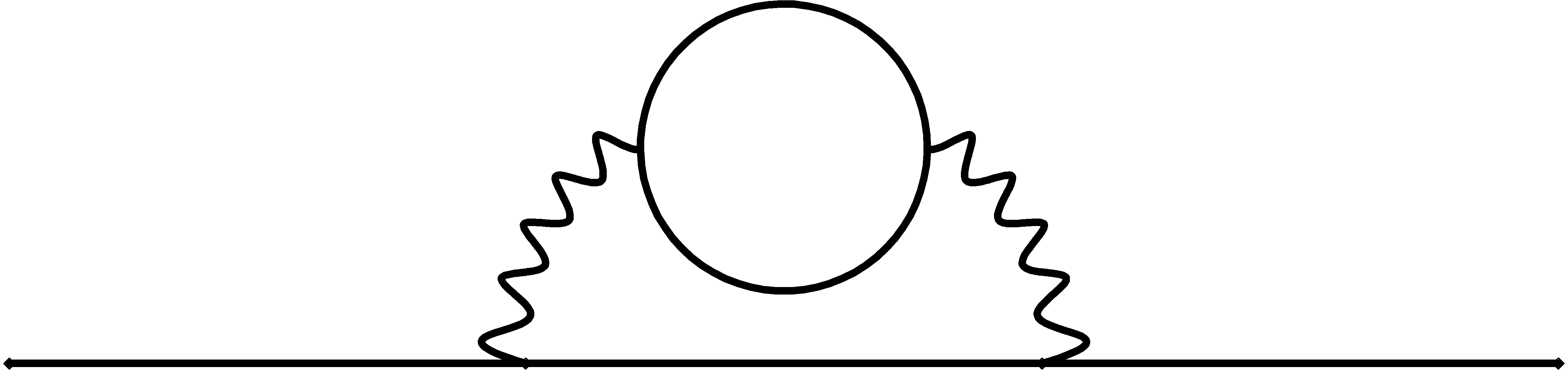}}
\vspace{+1ex}
\caption{Two-loop diagrams of such a form contain an additional power of growth with $ T $. In the Schwinger-Keldysh technique one has a sum of diagrams of such a form with vertexes attributed to different sides of the Keldysh's ``closed'' time contour.}
\label{image_dia_photon}
\end{figure}
The correction can be represented as:
\begin{equation}\label{2-loop-current-integral}
\Delta j_1(T) = 2e\int\dfrac{d^3\mathbf{p}}{(2\pi)^3}\Big(p_1+eA_1(T)\Big)\Bigg[
\left|f_\mathbf{p}(T)\right|^2 n_\mathbf{p}(T)+
\left(f_\mathbf{p}(T)\right)^2 \kappa_\mathbf{p}(T)+
\left(f^*_\mathbf{p}(T)\right)^2 \kappa^*_\mathbf{p}(T)
\Bigg],
\end{equation}
where the charged particle's level population and anomalous quantum average are equal to 
\begin{equation}
n_\mathbf{p}(T)=n^+_\mathbf{p}(T)+n^-_\mathbf{p}(T),\quad
(2\pi)^3 n^+_\mathbf{p}\delta\left(\mathbf{p}-\mathbf{p}'\right) =
\left\langle \widehat{b}_{-\mathbf{p}}^\dagger
\widehat{b}_{-\mathbf{p}'}\right\rangle,\quad 
(2\pi)^3 n^-_\mathbf{p}\delta\left(\mathbf{p}-\mathbf{p}'\right) =
\left\langle \widehat{a}_{\mathbf{p}}^\dagger
\widehat{a}_{\mathbf{p}'}\right\rangle
\end{equation}
and
\begin{equation}
(2\pi)^3\kappa_\mathbf{p}\delta\left(\mathbf{p}-\mathbf{p}'\right) =
\left\langle \widehat{a}_{\mathbf{p}}
\widehat{b}_{-\mathbf{p}'}\right\rangle.
\end{equation}
The integral expressions for the quantum expectation values are as follows:
\begin{equation}\label{n_p(T)-2-loop-normal}
\begin{split}
n^+_\mathbf{p}(T) &\simeq e^2\int\limits_{-T}^{T}d\tau_1 \int\limits_{-T}^{T}d\tau_2
\int\dfrac{d^3\mathbf{q}}{(2\pi)^3}\Delta G^K_{\mu\nu}(\mathbf{q};\tau_1,\tau_2)
f_{\mathbf{p}}(\tau_1)\overleftrightarrow{D_\mu}^\dagger
f_{\mathbf{p}-\mathbf{q}}(\tau_1)\cdot
f^*_{\mathbf{p}}(\tau_2)\overleftrightarrow{D_\nu}
f^*_{\mathbf{p}-\mathbf{q}}(\tau_2)
\end{split}
\end{equation}
and
\begin{equation}\label{n_p(T)-2-loop-amomal}
\begin{split}
\kappa_\mathbf{p}(T) &\simeq -e^2\int\limits_{-T}^{T}d\tau_1 \int\limits_{-T}^{\tau_1}d\tau_2
\int\dfrac{d^3\mathbf{q}}{(2\pi)^3}\Delta G^K_{\mu\nu}(\mathbf{q};\tau_1,\tau_2)
f^*_{\mathbf{p}}(\tau_1)\overleftrightarrow{D_\mu}^\dagger
f_{\mathbf{p}-\mathbf{q}}(\tau_1)\cdot
f^*_{\mathbf{p}}(\tau_2)\overleftrightarrow{D_\nu}
f^*_{\mathbf{p}-\mathbf{q}}(\tau_2) \\
&- e^2\int\limits_{-T}^{T}d\tau_1 \int\limits_{-T}^{\tau_1}d\tau_2
\int\dfrac{d^3\mathbf{q}}{(2\pi)^3}\Delta G^K_{\mu\nu}(-\mathbf{q};\tau_1,\tau_2)
f^*_{\mathbf{p}}(\tau_1)\overleftrightarrow{D_\mu}
f_{\mathbf{p}-\mathbf{q}}(\tau_1)\cdot
f^*_{\mathbf{p}}(\tau_2)\overleftrightarrow{D_\nu}^\dagger
f^*_{\mathbf{p}-\mathbf{q}}(\tau_2).
\end{split}
\end{equation}
The expression for $ n^-_\mathbf{p}(T) $ can be obtained from (\ref{n_p(T)-2-loop-normal}) by changing
\begin{equation}
\overleftrightarrow{D}_\mu^\dagger \mapsto\overleftrightarrow{D}_\mu,\quad
\overleftrightarrow{D}_\nu \mapsto\overleftrightarrow{D}_\nu^\dagger,\quad
\Delta G^K_{\mu\nu}(\mathbf{q};\tau_1,\tau_2)\mapsto \Delta G^K_{\mu\nu}(-\mathbf{q};\tau_1,\tau_2).
\end{equation}

Since the integrations over $ |\tau_{1,2}|$ are performed within the duration of the pulse, in the leading order in $ T $ we can use the WKB approximation for the modes as if the background field is eternal, i.e., $ A_1(\tau_i) \simeq E\tau_i $. Then, the modes can be approximately represented as $ f_\mathbf{k}(\tau_i)\simeq g_\mathbf{k}(k_1+eE\tau_i) $, where
\begin{equation}\label{WKB_Linear}
g_\mathbf{k}(K) \simeq
\dfrac{C_+(\mathbf{k},K)}{\sqrt{2\omega_{\mathbf{k}_\perp}(K)}}
\exp\left[\dfrac{i}{eE}\int\limits_{0}^{K}\omega_{\mathbf{k}_\perp}(y)dy \right]+
\dfrac{C_-(\mathbf{k},K)}{\sqrt{2\omega_{\mathbf{k}_\perp}(K)}}
\exp\left[-\dfrac{i}{eE}\int\limits_{0}^{K}\omega_{\mathbf{k}_\perp}(y)dy \right],
\end{equation} 
and
\begin{equation}
\omega_{\mathbf{k}_\perp}(K) = \sqrt{K^2+\mathbf{k}^2_\perp+m^2}, \quad K = k_1 + eE\tau,
\end{equation}
we will refer to these harmonic functions as $g$-modes as opposed to the $f$-modes from (\ref{WKB-General}).
Here, the moduli of the coefficients $ C_\pm(\mathbf{k},K) $ are defined in (\ref{WKB_C_refined}) and (\ref{WKB_C_mod}), while the phases are fixed by setting $ \tau_0=-k_1/(eE) $ in (\ref{WKB-General}). The integral over $ y $ in the exponents in (\ref{WKB_Linear}) can be calculated exactly,
\begin{equation}\label{WKB-integral-2}
\begin{split}
\dfrac{1}{eE}\int\limits_{0}^{K}\omega_{\mathbf{k}_\perp}(y)dy &=
\dfrac{1}{eE}\int\limits_{0}^{K} \sqrt{y^2+\mathbf{k}^2_\perp+m^2}dy = \\
&= \dfrac{K}{2eE}\sqrt{K^2+\mathbf{k}^2_\perp+m^2} + \\
&+ \sgn(K)\dfrac{\mathbf{k}^2_\perp+m^2}{2eE}\ln\left[
\sqrt{1+\dfrac{K^2}{\mathbf{k}^2_\perp+m^2}}+
\dfrac{|K|}{\sqrt{\mathbf{k}^2_\perp+m^2}}
\right],
\end{split}
\end{equation}
and then conveniently approximated for different values of $ K $. Namely, for $ |K|<\sqrt{\mathbf{k}^2_\perp+m^2} $, we can expand (\ref{WKB-integral-2}) in the neighborhood of $ K=0 $ to obtain:
\begin{equation}\label{small-arg}
\dfrac{1}{eE}\int\limits_{0}^{K}\omega_{\mathbf{k}_\perp}(y)dy=\dfrac{K\sqrt{\mathbf{k}^2_\perp+m^2}}{eE}\left[1+\mathcal{O}\left[
\dfrac{K^2}{\mathbf{k}^2_\perp+m^2}\right]\right].
\end{equation}
At the same time, for $ |K|>\sqrt{\mathbf{k}^2_\perp+m^2} $ we find the following asymptotic form of the modes:
\begin{equation}\label{cylinder_asymptotics-2}
\begin{split}
g_\mathbf{k}(K) &\simeq
\dfrac{C_{\sgn(K)}(\mathbf{k},K)}{\sqrt{2\sqrt{eE}}}
\left|\dfrac{K}{\sqrt{eE}}\right|^{i\frac{m^2+\mathbf{k}^2_\perp}{2eE}-\frac{1}{2}}
\exp\left[\dfrac{iK^2}{2eE}\right] + \\
&+ \dfrac{C_{-\sgn(K)}(\mathbf{k},K)}{\sqrt{2\sqrt{eE}}}
\left|\dfrac{K}{\sqrt{eE}}\right|^{-i\frac{m^2+\mathbf{k}^2_\perp}{2eE}-\frac{1}{2}}\exp\left[-\dfrac{iK^2}{2eE}\right].
\end{split}
\end{equation}
For all loop integrals, which we consider in this paper, we will estimate only the contribution to them that come from the integrations over the regions, where the asymptotic form (\ref{cylinder_asymptotics-2}) is applicable for each $ g $-mode in the expression. In Appendix \ref{appendix-subleading}, we show that the contribution from the narrow complementary region of integration is suppressed compared to the leading contributions that we find here.

\section{One-loop corrections to the photon's propagator in the limit of soft momenta}

In this subsection, we estimate the one-loop correction to the matrix propagator of photons, $ \Delta G^{ab}_{\mu\nu}(\mathbf{q};\tau_1,\tau_2),\\ a,b=\pm $, for $ |\mathbf{q}|=\mathcal{O}(1/T) $ and $ |\tau_{1,2}|\lesssim T $, where $ \tau_{1,2} $ do not have to obey the condition (\ref{Thelimit}). It is convenient at this point to rewrite the loop correction $ \Delta G_{\mu\nu} $ in terms of advanced, retarded, and Keldysh propagators:
\begin{equation}
G^R=G^{--}-G^{-+},\quad
G^A=G^{--}-G^{+-},\quad
G^K=\dfrac{1}{2}\Big[ G^{-+}+G^{+-} \Big].
\end{equation}
Then in the Schwinger-Keldysh technique, the self-energy $ \Sigma_{\mu\nu} $ is the matrix of the form
\begin{equation}
\begin{pmatrix}
2\Sigma^K & \Sigma^R \\
\Sigma^A & 0
\end{pmatrix}=\overleftrightarrow{D}_\mu \overleftrightarrow{D}_\nu^\dagger
\begin{pmatrix}
2D_\mathbf{k}^KD_{\mathbf{k}-\mathbf{q}}^K+\frac{1}{2}\left[D_\mathbf{k}^AD_{\mathbf{k}-\mathbf{q}}^A+D_\mathbf{k}^RD_{\mathbf{k}-\mathbf{q}}^R\right] & D_\mathbf{k}^KD_{\mathbf{k}-\mathbf{q}}^R+D_\mathbf{k}^RD_{\mathbf{k}-\mathbf{q}}^K \\
D_\mathbf{k}^KD_{\mathbf{k}-\mathbf{q}}^A+D_\mathbf{k}^AD_{\mathbf{k}-\mathbf{q}}^K & 0
\end{pmatrix},
\end{equation} 
where $ D_\mathbf{k}^{K,R,A}=D^{K,R,A}(\mathbf{k};\tau_3,\tau_4) $ are the tree-level Keldysh, retarded and advanced propagators of the scalar field. This, we have the following general expressions for the components of the self-energy,
\begin{equation}
\Sigma_{\mu\nu}^{A,R}\propto f_{\mathbf{k}}\left(\tau_3\right) \overleftrightarrow{D}_\mu f_{\mathbf{k}-\mathbf{q}}\left(\tau_3\right) \cdot f_{\mathbf{k}}^*\left(\tau_4\right) \overleftrightarrow{D}_\nu^{\dagger} f_{\mathbf{k}-\mathbf{q}}^*\left(\tau_4\right)-
f^*_{\mathbf{k}}\left(\tau_3\right) \overleftrightarrow{D}_\mu f^*_{\mathbf{k}-\mathbf{q}}\left(\tau_3\right) \cdot f_{\mathbf{k}}\left(\tau_4\right) \overleftrightarrow{D}_\nu^{\dagger} f_{\mathbf{k}-\mathbf{q}}\left(\tau_4\right),
\end{equation}
and
\begin{equation}
\begin{split}
\Sigma_{\mu\nu}^{K}(\mathbf{k},\mathbf{q};\tau_3,\tau_4) &= \theta(\tau_1-\tau_3)\theta(\tau_2-\tau_4)\times \\
&\times\left[
f_{\mathbf{k}}\left(\tau_3\right) \overleftrightarrow{D}_\mu f_{\mathbf{k}-\mathbf{q}}\left(\tau_3\right) \cdot f_{\mathbf{k}}^*\left(\tau_4\right) \overleftrightarrow{D}_\nu^{\dagger} f_{\mathbf{k}-\mathbf{q}}^*\left(\tau_4\right)+ \right. \\
&+\left.
f^*_{\mathbf{k}}\left(\tau_3\right) \overleftrightarrow{D}_\mu f^*_{\mathbf{k}-\mathbf{q}}\left(\tau_3\right) \cdot f_{\mathbf{k}}\left(\tau_4\right) \overleftrightarrow{D}_\nu^{\dagger} f_{\mathbf{k}-\mathbf{q}}\left(\tau_4\right)\right].
\end{split}
\end{equation}
Since $ \tau_{3,4} $ lie within the duration of the pulse, we use the expression for the $ g $-modes (the WKB approximation of the modes in the constant eternal background field with the appropriate initial conditions at the beginning of the pulse) from (\ref{WKB_Linear}) rather than the $f$-modes (the WKB approximation for the modes in the pulse background). Using the asymptotic form (\ref{cylinder_asymptotics-2}) for each mode in $ \Sigma^{A,R,K}_{\mu\nu} $ and extracting the leading terms (one can see details of the calculation in Appendix \ref{appendix-details}), one finds straightforwardly that $ \Sigma^{A,R}_{\mu\nu} $ are suppressed by $ |\mathbf{q}|\sim 1/T $ compared to $ \Sigma^K_{\mu\nu} $. As a result, we can neglect the corrections to the advanced and retarded propagators compared to the correction to the Keldysh one, which is a typical situation \cite{kamenev2023field}. The expression for one-loop correction to the Keldysh propagator $ \Delta G^K_{\mu\nu}(\mathbf{q};\tau_1,\tau_2) $ is as follows:
\begin{equation}\label{Delta-G-K-soft-int}
\Delta G^K_{\mu\nu}(\mathbf{q};\tau_1,\tau_2) \simeq e^2
\int d\tau_3 \int d\tau_4 \int\dfrac{d^3\mathbf{k}}{(2\pi)^3}
G^R(\mathbf{q};\tau_1,\tau_3)\Sigma^K_{\mu\nu}(\mathbf{k},\mathbf{q};\tau_3,\tau_4)G^A(\mathbf{q};\tau_4,\tau_2),
\end{equation}
where we use that the tree-level propagators $ G^{R,A}_{\mu\nu} $ are proportional to the metric tensor $ g_{\mu\nu} $, i.e. $ G^{R,A}_{\mu\nu}=g_{\mu\nu}G^{R,A} $.

Applying the asymptotic form (\ref{cylinder_asymptotics-2}) to each $g$-mode in $ \Sigma^K_{\mu\nu}(\mathbf{k},\mathbf{q};\tau_3,\tau_4) $ and singling out such contributions, which have the highest powers of $ T $ and do not oscillate with $ \mathbf{q} $, we obtain (see Appendix \ref{appendix-details}) such expression for $ \Delta G^K_{\mu\nu}(\mathbf{q};\tau_1,\tau_2) $, which gives the leading contribution to the two-loop correction to the current: 
\begin{equation}\label{Delta-G-K-soft}
\begin{split}
\Delta G^K_{\mu\nu}(\mathbf{q};\tau_1,\tau_2) &\simeq \sum\limits_{\alpha,\alpha_0=-}^{+} e^2 \int\dfrac{d^3\mathbf{k}}{(2\pi)^3}
\int\limits_{-k_1/(eE)}^{\tau_1}d\tau_3 
\int\limits_{-k_1/(eE)}^{\tau_2}d\tau_4 \;
|C_-(\mathbf{k},+\infty)|^2 |C_+(\mathbf{k},+\infty)|^2 \\
&\times \left[ e^{2i\alpha q_1(\tau_3-\tau_4)}
\begin{pmatrix}
1 & \alpha \\
\alpha & 1
\end{pmatrix}_{\mu\nu} + e^{2i\alpha q_1(2k_1/(eE)+\tau_3+\tau_4)}
\begin{pmatrix}
-1 & \alpha \\
-\alpha & 1
\end{pmatrix}_{\mu\nu} \right]\cdot
\dfrac{e^{i\alpha_0|\mathbf{q}|[(\tau_1-\tau_2)-(\tau_3-\tau_4)]}}{4|\mathbf{q}|^2}.
\end{split}
\end{equation}
Components of the photon's Keldysh propagator with $\mu, \nu = \overline{2,3}$ receive suppressed contributions compared to the components with $\mu,\nu = \overline{0,1}$, which are present in this expression.

It should be emphasized that the obtained one-loop correction to the propagator does not contain UV divergences, because we consider propagator for which only a partial Fourier transformation over the spatial coordinates has been performed. In this case, there are insufficient integrations to produce UV divergences. To reveal UV divergences, we would need either to perform an inverse Fourier transformation to $ \mathbf{x} $-space or a full Fourier transformation over all space-time directions, if possible. 

\section{Two-loop correction to the electric current}

In this subsection, we estimate the leading two-loop correction to the current by the end of the pulse, $t \lesssim T$. According to (\ref{2-loop-current-integral}) to do that we should calculate $n^\pm_\mathbf{p}(T)$, $ \kappa_\mathbf{p}(T) $ and $ \kappa^*_\mathbf{p}(T) $, where

\begin{equation}\label{n_p(T)-2-loop-normal-g}
\begin{split}
n^+_\mathbf{p}(T) &\simeq e^2\int\limits_{-T}^{T}d\tau_1 \int\limits_{-T}^{T}d\tau_2
\int\dfrac{d^3\mathbf{q}}{(2\pi)^3} \Delta G^K_{\mu\nu}(\mathbf{q};\tau_1,\tau_2) \times \\
&\times
g_{\mathbf{p}}(p_1+eE\tau_1)\overleftrightarrow{D_\mu}^\dagger
g_{\mathbf{p}-\mathbf{q}}(p_1-q_1+eE\tau_1)\cdot
g^*_{\mathbf{p}}(p_1+eE\tau_2)\overleftrightarrow{D_\nu}
g^*_{\mathbf{p}-\mathbf{q}}(p_1-q_1+eE\tau_2),
\end{split}
\end{equation}
while $ \Delta G^K_{\mu\nu}(\mathbf{q};\tau_1.\tau_2) $ has been calculated above. For the anomalous expectation value, $ \kappa_\mathbf{p}(T) $, there is an expression, which is similar to (\ref{n_p(T)-2-loop-normal-g}), the only difference being that integration over $ \tau_2 $ is performed here along the interval $ -T<\tau_2<\tau_1 $, and we should make the substitution $ g_{\mathbf{p}}(p_1+eE\tau_1)\mapsto-g^*_{\mathbf{p}}(p_1+eE\tau_1) $. 

After the substitution of the asymptotic form (\ref{cylinder_asymptotics-2}) into the integral (\ref{n_p(T)-2-loop-normal-g}), we find that $ n^-_\mathbf{p}(T)\simeq n^+_\mathbf{p}(T) = n_\mathbf{p}(T)$ and the integral is saturated around $ |\mathbf{q}|=\mathcal{O}(1/T) $. As a result (see the Appendix \ref{2-loop-Appendix} for more details), we obtain the leading in $ T $ expression for $ n_\mathbf{p}(T) $:
\begin{equation}\label{n-p(T)-leading}
\begin{split}
n_\mathbf{p}(T) &\simeq \dfrac{4e^4}{\pi}\sum\limits_{\alpha=-}^{+}
\int\limits_{-p_1/{eE}}^{T}d\tau_1 \int\limits_{-p_1/{eE}}^{T}d\tau_2
\int\dfrac{d^3\mathbf{k}}{(2\pi)^3}
\int\limits_{-\frac{k_1}{eE}+|\tau|}^{\mathcal{T}}d\mathcal{T}_{34}
\\
&\times
|C_+(\mathbf{k},+\infty)|^2 |C_-(\mathbf{k},+\infty)|^2
|C_\alpha(\mathbf{p},+\infty)|^2 |C_{-\alpha}(\mathbf{p},+\infty)|^2,
\end{split} 
\end{equation}
where
\begin{equation}\label{varchange-2}
\tau=\frac{\tau_1-\tau_2}{2}, \quad \mathcal{T}=\frac{\tau_1+\tau_2}{2}, \quad
\mathcal{T}_{34}=\frac{\tau_3+\tau_4}{2}.
\end{equation}
To calculate $ \kappa_\mathbf{p}(T) $ we should keep in mind that,
as follows from (\ref{cylinder_asymptotics-2}), in the asymptotic region $ g^*_{\mathbf{p}}(p_1+eE\tau_1) $ differs from $ g_{\mathbf{p}}(p_1+eE\tau_1) $ only by the coefficients $ C_\pm $. This leads to the following change in (\ref{n-p(T)-leading}):
\begin{equation}\label{anomal-0.0}
|C_\alpha(\mathbf{p},+\infty)|^2 \mapsto - 
C^*_{-\alpha}(\mathbf{p},+\infty)
C^*_\alpha(\mathbf{p},+\infty).
\end{equation}
Then, we obtain the following expression,
\begin{equation}\label{kappa-p(T)-leading}
\begin{split}
\kappa_\mathbf{p}(T) &\simeq -\dfrac{4e^4}{\pi}\sum\limits_{\alpha=-}^{+}
\int\limits_{-p_1/{eE}}^{T}d\tau_1 \int\limits_{-p_1/{eE}}^{\tau_1}d\tau_2
\int\dfrac{d^3\mathbf{k}}{(2\pi)^3}
\int\limits_{-\frac{k_1}{eE}+|\tau|}^{\mathcal{T}}d\mathcal{T}_{34}
\\
&\times
|C_+(\mathbf{k},+\infty)|^2 |C_-(\mathbf{k},+\infty)|^2 \\
&\times
C^*_{-\alpha}(\mathbf{p},+\infty)
C^*_\alpha(\mathbf{p},+\infty) |C_{-\alpha}(\mathbf{p},+\infty)|^2.
\end{split} 
\end{equation}
Within the correction for the current (\ref{2-loop-current-integral}), the level-population $ n_\mathbf{p}(T) $ is multiplied by $ |f_\mathbf{p}(T)|^2 $, while the anomalous expectation value $ \kappa_\mathbf{p}(T) $ is multiplied by $ (f_\mathbf{p}(T))^2$, where
\begin{equation}\label{anomal-0.1}
\begin{split}
|f_\mathbf{p}(T)|^2 &\simeq 
\dfrac{1+2|C_+(\mathbf{p},+\infty)|^2}
{2\omega_\mathbf{p}(T)}+
\sum\limits_{\beta=-}^{+}
\dfrac{C_\beta(\mathbf{p},+\infty)C^*_{-\beta}(\mathbf{p},+\infty)}{2\omega_\mathbf{p}(T)}
\cdot\exp\left[ 2i\beta \int\limits_{t_0}^{T} \omega_\mathbf{p}(\tau) d\tau \right], \\
(f_\mathbf{p}(T))^2 &\simeq 
\dfrac{2C_+(\mathbf{p},+\infty)C_-(\mathbf{p},+\infty)}
{2\omega_\mathbf{p}(T)}+
\sum\limits_{\beta=-}^{+}
\dfrac{C^2_\beta(\mathbf{p},+\infty)}{2\omega_\mathbf{p}(T)}
\cdot\exp\left[ 2i\beta \int\limits_{t_0}^{T} \omega_\mathbf{p}(\tau) d\tau \right].
\end{split}
\end{equation}
There is also the contribution from $ \kappa^*_\mathbf{p}(T) $ in (\ref{2-loop-current-integral}), which is complex conjugate to $ \kappa_\mathbf{p}(T) $ and should be taken into account.
Substituting (\ref{n-p(T)-leading} - \ref{anomal-0.1}) in (\ref{2-loop-current-integral}), we obtain the leading two-loop contribution to the electric current:
\begin{equation}
\begin{split}
\Delta j_1(T) &\simeq -2e\int\dfrac{d^3\mathbf{p}}{(2\pi)^3}\dfrac{p_1+eA_1(T)}{2\omega_\mathbf{p}(T)}\cdot\dfrac{4e^4}{\pi}
\int\limits_{-p_1/{eE}}^{T}d\tau_1 \int\limits_{-p_1/{eE}}^{T}d\tau_2 \times \\
&\times
\int\dfrac{d^3\mathbf{k}}{(2\pi)^3}
\int\limits_{-\frac{k_1}{eE}+|\tau|}^{\mathcal{T}}d\mathcal{T}_{34}
|C_+(\mathbf{k},+\infty)|^2 |C_-(\mathbf{k},+\infty)|^2 \times \\
&\times \dfrac{1}{2}\left[ 
C_+(\mathbf{p},+\infty)C^*_-(\mathbf{p},+\infty)
\cdot\exp\left[ 2i\int\limits_{t_0}^{T} \omega_\mathbf{p}(\tau) d\tau \right]
+\text{c.c.}.
\right]. 
\end{split}
\end{equation}
Since $ |k_1|/(eE),|p_1|/(eE)\sim T/\mu\ll T $, we can neglect $ p_1 $ and $ k_1 $ in the limits of the integrals over $ \tau_{1,2} $ and $ \mathcal{T}_{34} $, correspondingly, to obtain that
\begin{equation}\label{2-loop-current-leading-pre-final}
\begin{split}
\Delta j_1(T) &\simeq -2e\int\dfrac{d^3\mathbf{p}}{(2\pi)^3}\dfrac{p_1+eA_1(T)}{2\omega_\mathbf{p}(T)}\cdot\dfrac{2e^4}{\pi}
\int\limits_{0}^{T}d\tau_1 \int\limits_{0}^{T}d\tau_2
\int\dfrac{d^3\mathbf{k}}{(2\pi)^3}
\int\limits_{|\tau|}^{\mathcal{T}}d\mathcal{T}_{34}
|C_+(\mathbf{k},+\infty)|^2\times \\
&\times \left[ 
C_+(\mathbf{p},+\infty)C^*_-(\mathbf{p},+\infty)
\cdot\exp\left[ 2i\int\limits_{t_0}^{T} \omega_\mathbf{p}(\tau) d\tau \right]
+\text{c.c.}
\right] = \\
&= -j_1^{(\text{tree,osc})}(T) \dfrac{2e^4}{\pi}
\int\limits_{0}^{T}d\tau_1 \int\limits_{0}^{T}d\tau_2
\int\limits_{|\tau|}^{\mathcal{T}}d\mathcal{T}_{34}
\int\dfrac{d^3\mathbf{k}}{(2\pi)^3}
|C_+(\mathbf{k},+\infty)|^2,
\end{split}
\end{equation}
where $ j_1^{(\text{tree,osc})} $ is given by (\ref{delta-j-tree}) with $t \simeq T$.
After the integration over $ \tau_1,\tau_2,\mathcal{T}_{34} $ and $ \mathbf{k} $, we obtain the final expression for the leading two-loop correction to the electric current,
\begin{equation}\label{2-loop-current-leading}
\boxed{\Delta j_1(T) = -j_1^{(\text{tree,osc})}(T)\dfrac{2e^4}{\pi}\cdot\dfrac{T^3}{3}\cdot\dfrac{e^2 E^2 T}{8\pi^3\mu}
\exp\left[-\dfrac{\pi m^2}{eE}\right].}
\end{equation}
We can see that non-oscillating contributions from (\ref{anomal-0.1}) cancel each other\footnote{A somewhat similar situation is encountered in the calculation of loop corrections for light fields in de Sitter spacetime \cite{Akhmedov:2017ooy,Akhmedov:2024npw}. There, leading power-like corrections (in conformal time) to level populations and anomalous expectation values cancel each other within the Keldysh propagator, leaving only logarithmic corrections.} within $\Delta j_1(T)$. Consequently, the leading result (\ref{2-loop-current-leading}) for the two-loop correction, in contrast to the tree-level expression, consists solely of oscillating terms. However, as the function of the pulse duration $T$ this correction grows much faster than the tree-level current, $T^4$ versus $T$, so that for sufficiently large $T$, the correction becomes dominant.

In Appendix \ref{appendix-subleading}, we show that all subleading contributions to the second-loop correction to the current, which we neglected during our calculations, are of the order $ \mathcal{O}(T^3) $ or lower.

\section{Conclusions and acknowledgments}

Thus, for a very long pulse with $eET \gg m$, the leading contribution to the tree-level current is given by eq. (\ref{tree_current_renorm_finally_weak}). Meanwhile, the relevant correction for the long pulse is given by eq. (\ref{2-loop-current-leading}). For $t \lesssim T$, this correction contains the second power of the Schwinger exponent $\exp\left[-\pi m^2/eE\right]$, i.e., one extra power compared to the tree-level current. However, it grows as $T^4$, which is much faster than the growth of the tree-level current, $T^1$. Hence, for a sufficiently long pulse, the correction can dominate over the leading contribution.

Of course, for currently achievable electric field strengths, the Schwinger exponent is very small, and the effects considered here are negligible. However, for fields on the order of the Schwinger critical value, there is no suppression by the exponent, and the effects discussed here may play an essential role. In that case, one must perform a resummation of the leading corrections from all loops, which remains a subject for future study.

We would like to acknowledge discussions with A. Semenov. This work was partially supported by the Ministry of Science and Higher Education of the Russian Federation (agreement no. 075–15–2022–287).

\appendix

\section{Details of the loop integrals calculation}
\subsection{Calculation of one-loop correction to the photon's propagator in the limit of soft momenta}\label{appendix-details}

In this subsection, we provide detailed calculations that lead to the expression (\ref{Delta-G-K-soft}). In previous sections, we already noticed that we should use the expression (\ref{WKB_Linear}) for the $ g $-modes rather than the exact modes $f$, since $ \tau_{3,4} $ are within the duration of the pulse. Then, we apply the asymptotic form (\ref{cylinder_asymptotics-2}) for each $g$-mode and obtain the following expression for $ \Delta G^K_{\mu\nu}(\mathbf{q};\tau_1,\tau_2) $,
\begin{equation}\label{Delta-G-K-soft-2}
\begin{split}
\Delta G^K_{\mu\nu}(\mathbf{q};\tau_1,\tau_2) &\simeq e^2\int d\tau_3\int d\tau_4 \int\dfrac{d^3\mathbf{k}}{(2\pi)^3} \Sigma^K_{\mu\nu}(\mathbf{k},\mathbf{q};\tau_3,\tau_4) \\
&\times\sum\limits_{\alpha_0=-}^{+}
\left[\dfrac{e^{i \alpha_0|\mathbf{q}|((\tau_1-\tau_2)-(\tau_3-\tau_4))}}{4|\mathbf{q}|^2}-\frac{e^{i \alpha_0|\mathbf{q}|((\tau_1+\tau_2)-(\tau_3+\tau_4))}}{4|\mathbf{q}|^2}\right],
\end{split}
\end{equation}
where
\begin{equation}\label{Sigma-K-int}
\begin{split}
\Sigma^K_{\mu\nu}(\mathbf{k},\mathbf{q};\tau_3,\tau_4) \simeq \sum\limits_{\alpha,\beta,\gamma,\delta=-}^{+}
\dfrac{\theta(\tau_1-\tau_3)\theta(\tau_2-\tau_4)}{4eE}
\cdot\overleftrightarrow{D}_\mu\cdot \overleftrightarrow{D}_\nu^\dagger\\
\times \left|\dfrac{k_1+eE\tau_3}{\sqrt{eE}}\right|^{i(\alpha+\beta)\frac{m^2+\mathbf{k}^2_\perp}{2eE}-1}
\left|\dfrac{k_1+eE\tau_4}
{\sqrt{eE}}\right|^{-i(\gamma+\delta)\frac{m^2+\mathbf{k}^2_\perp}{2eE}-1} \\
\times\exp\left[
\dfrac{i\alpha(k_1+eE\tau_3)^2}{2eE}+\dfrac{i\beta(k_1-q_1+eE\tau_3)^2}{2eE}
-\dfrac{i\gamma(k_1+eE\tau_4)^2}{2eE}-\dfrac{i\delta(k_1-q_1+eE\tau_4)^2}{2eE}
\right] \\
\times\bigg[
C_{\alpha\sgn(k_1+eE\tau_3)}(\mathbf{k},k_1+eE\tau_3)
C_{\beta\sgn(k_1+eE\tau_3)}(\mathbf{k},k_1+eE\tau_3) \\
\times C^*_{\gamma\sgn(k_1+eE\tau_4)}(\mathbf{k},k_1+eE\tau_4)
C^*_{\delta\sgn(k_1+eE\tau_4)}(\mathbf{k},k_1+eE\tau_4) + \\ 
+ C^*_{-\alpha\sgn(k_1+eE\tau_3)}(\mathbf{k},k_1+eE\tau_3)
C^*_{-\beta\sgn(k_1+eE\tau_3)}(\mathbf{k},k_1+eE\tau_3) \\
\times \ C_{-\gamma\sgn(k_1+eE\tau_4)}(\mathbf{k},k_1+eE\tau_4)
C_{-\delta\sgn(k_1+eE\tau_4)}(\mathbf{k},k_1+eE\tau_4)
\bigg]
\end{split}
\end{equation}
and
\begin{equation}\label{D-from-vertex-soft}
\begin{split}
\overleftrightarrow{D}_{\mu\neq 0}=i(2k_\mu+2eE\tau_3\delta_{1\mu}),&\quad
\overleftrightarrow{D}_{\nu\neq 0}^\dagger=-i(2k_\nu+2eE\tau_4\delta_{1\nu});
\\
\overleftrightarrow{D}_{\mu=0}=i(\alpha-\beta)(k_1+eE\tau_3),&\quad
\overleftrightarrow{D}_{\nu=0}^\dagger=-i(\gamma-\delta)(k_1+eE\tau_4).
\end{split}
\end{equation}
Since $ |\mathbf{q}|=\mathcal{O}(1/T) $, we neglected $ \mathbf{q} $ in the pre-exponent of (\ref{Sigma-K-int}). For the same reason we can neglect $ q_1^2 $ in the exponent. During the calculation of the second loop correction to the current, we integrate over $ |\mathbf{q}| $. Thus, in the corresponding integrand we should keep only those terms that do not rapidly oscillate with $ |\mathbf{q}| $; otherwise, the integral will be suppressed. From (\ref{Sigma-K-int}), we see that $ \tau_1+\tau_2>\tau_3-\tau_4 $, which implies that the second exponent in the last line in eq. (\ref{Delta-G-K-soft-2}) does not oscillate only in a small neighborhood of the point $ (\tau_3=\tau_1,\tau_4=\tau_2) $, while the first exponent does not oscillate on the line $ \{\tau_3-\tau_4 = \tau_1-\tau_2 \} $. Thus, the contribution of the exponent $ \exp(i\alpha_0|\mathbf{q}|[(\tau_1+\tau_2)-(\tau_3+\tau_4)]) $ can be neglected\footnote{Rephrasing this, we keep only those contributions to (\ref{Delta-G-K-soft-2}) that do not oscillate with respect to $ \mathbf{q} $, since the expression for the photon propagator will be used to calculate the two-loop correction to the electric current, which contains the integral over $ \mathbf{q} $.}.

On the other hand, (\ref{Sigma-K-int}) contains rapidly oscillating exponents of $ k_1^2 $ or $ \tau^2_{1,2} $ for generic values of $ \alpha, \beta, \gamma, \delta = \pm $. These oscillations suppress the integrals over $ k_1,\tau_{1,2} $. Since, we are interested only in the contributions in (\ref{Delta-G-K-soft-2}) which provide the leading (in $ T $) corrections to the current in the second loop, we should consider only those values of $ \alpha, \beta, \gamma, \delta $, for which there are no such oscillating exponents. As a result, we can keep only two terms, $ \alpha=-\beta=\pm\gamma=\mp\delta $.

One can see that for $ \alpha=-\beta=\pm\gamma=\mp\delta $, the terms in the bracket in (\ref{Sigma-K-int}) are the same and equal to $ \theta(k_1+eE\tau_3)\theta(k_1+eE\tau_4)|C_-(\mathbf{k},+\infty)|^2 |C_+(\mathbf{k},+\infty)|^2 $. For $ \Sigma^{AR}_{\mu\nu} $, we can provide similar calculations and obtain an expression of the type (\ref{Sigma-K-int}) but with the difference inside the bracket, rather than the sum. For $ \alpha=-\beta=\pm\gamma=\mp\delta $, this difference is suppressed in $ |\mathbf{q}|=\mathcal{O}(1/T) $. As a result, both $ \Sigma^{A,R}_{\mu\nu} $ are suppressed compared to the leading contribution of $ \Sigma^{K}_{\mu\nu} $. 

Finally, after the substitution $ \alpha=-\beta=\pm\gamma=\mp\delta $ into (\ref{Sigma-K-int}), we find that the contributions with $ \mu,\nu\in\{0,1\} $ appear to be the leading ones, and obtain the expression (\ref{Delta-G-K-soft}).

\subsection{Calculation of the two-loop correction to $ n_\mathbf{p}(T) $}\label{2-loop-Appendix}

Similarly to the derivation of the expression (\ref{Sigma-K-int}), we can substitute the asymptotics (\ref{cylinder_asymptotics-2}) into the integral (\ref{n_p(T)-2-loop-normal-g}) and obtain a sum of 16 terms over $ \alpha',\beta',\gamma',\delta'=\pm $. Substituting (\ref{Delta-G-K-soft-2}) and (\ref{Sigma-K-int}) in (\ref{n_p(T)-2-loop-normal-g}), we obtain a sum of 256 terms over $ \alpha,\beta,\gamma,\delta,\alpha',\beta',\gamma',\delta'=\pm $. After eliminating the terms that oscillate in $ p_1^2 $, $ k_1^2 $ and $ \tau^2_{1,2,3,4}$, we obtain the following conditions,
\begin{equation}\label{non-osc-options}
\alpha = -\beta=\pm\gamma=\mp\delta 
\quad\quad \text{and} \quad\quad
\alpha' = -\beta'=\pm\gamma'=\mp\delta'. 
\end{equation}
As a result, the integrand of $ n^+_\mathbf{p}(T) $ (which is a sum over $ \alpha=\pm $ and an integral over $ \mathbf{q}, \tau_{1,2,3,4} $ etc.) is proportional to\footnote{We extract the matrix structure and exponents, which depend on $ |\mathbf{q}| $ and/or $ \tau_{1,2,3,4} $, from the integrand.}
\begin{equation}
\begin{split}
\Psi &= e^{2i\alpha_0|\mathbf{q}|(\tau-\tau_{34})} \left[ 
e^{2i\alpha q_1\tau_{34}}
\begin{pmatrix}
1 & -\alpha \\
-\alpha & 1
\end{pmatrix}^{\mu\nu}+
e^{2i\alpha q_1(k_1/(eE)+\mathcal{T}_{34})}
\begin{pmatrix}
-1 & -\alpha \\
\alpha & 1
\end{pmatrix}^{\mu\nu}
\right] \\
&\times \left[
e^{2i\alpha' q_1\tau}
\begin{pmatrix}
1 & -\alpha' \\
-\alpha' & 1
\end{pmatrix}_{\mu\nu} +
e^{2i\alpha' q_1(p_1/(eE)+\mathcal{T})}
\begin{pmatrix}
-1 & -\alpha' \\
\alpha' & 1
\end{pmatrix}_{\mu\nu}\right],
\end{split}
\end{equation}
where $ \tau_{34}=(\tau_3-\tau_4)/2 $.

We see that the contraction over $ \mu,\nu $ is not suppressed only for $ \alpha'=\alpha $. We can also do the same for $ n^-_\mathbf{p}(T) $, which differs from $ n^+_\mathbf{p}(T) $ by changing $ \overleftrightarrow{D_\mu}^\dagger\overleftrightarrow{D_\nu}\mapsto \overleftrightarrow{D_\mu}\overleftrightarrow{D_\nu}^\dagger $ and $ \overleftrightarrow{D^\mu}\overleftrightarrow{D^\nu}^\dagger\mapsto
\overleftrightarrow{D^\mu}^\dagger\overleftrightarrow{D^\nu} $ for the phonon loop. However, it follows straightforwardly from (\ref{D-from-vertex-soft}) that such a change does not affect the contraction over $ \mu,\nu $. For this reason, $ n^-_\mathbf{p}(T)=n^+_\mathbf{p}(T) $. As a result, we have that:
\begin{equation}
\Psi = e^{2i\alpha_0|\mathbf{q}|(\tau-\tau_{34})} \left[
e^{2i\alpha q_1(\tau+\tau_{34})}+e^{2i\alpha q_1((p_1+k_1)/(eE)+\mathcal{T}+\mathcal{T}_{34})}
\right]
\end{equation}
It is convenient to represent the integral over $ \mathbf{q} $ as $ d^3\mathbf{q}=|\mathbf{q}|d|\mathbf{q}|dq_1 d\varphi $. In this case, we find that the integral over $ |\mathbf{q}| $ is saturated around $ \tau\simeq\tau_{34} $. Since $ (\tau-\tau_{34}) $ and $ (\tau+\tau_{34}) $ are linearly independent, we can integrate over $ (\tau-\tau_{34}) $ separately and obtain $ \delta(|\mathbf{q}|) $ in the leading order in $ T $. This is the reason why we considered soft photons with $ |\mathbf{q}|=\mathcal{O}(1/T) $ in the previous subsection.

Now, we can restore the leading contribution in $ n_\mathbf{p}(T) $:
\begin{equation}\label{2-loop-n-p(T)-leading}
\begin{split}
n_\mathbf{p}(T) &\simeq 2e^2 \sum\limits_{\alpha_0,\alpha=-}^{+}
\int\limits_{-p_1/{eE}}^{T}d\tau_1 \int\limits_{-p_1/{eE}}^{T}d\tau_2
\int\dfrac{d^3\mathbf{q}}{(2\pi)^3}\cdot e^2\int\dfrac{d^3\mathbf{k}}{(2\pi)^3}
\\
&\times 
\int\limits_{-k_1/(eE)}^{\tau_1}d\tau_3 \int\limits_{-k_1/(eE)}^{\tau_2}d\tau_4
\dfrac{e^{2i\alpha_0|\mathbf{q}|(\tau-\tau_{34})}}{4|\mathbf{q}|^2}
|C_+(\mathbf{k},+\infty)|^2 |C_-(\mathbf{k},+\infty)|^2
\\
&\times
|C_+(\mathbf{p},+\infty)|^2 |C_-(\mathbf{p},+\infty)|^2 \cdot 4
\left[e^{2i\alpha q_1(\tau+\tau_{34})} +
e^{2i\alpha q_1((p_1+k_1)/(eE)+\mathcal{T}+\mathcal{T}_{34})}\right].
\end{split}
\end{equation}
After the integration over $ (\tau-\tau_{34}) $ and $ \mathbf{q} $, we obtain (\ref{n-p(T)-leading}).

\section{Analysis of subleading contributions}\label{appendix-subleading}

In the main text we estimate contribution to loop integrals coming from the region of parameters, where (\ref{cylinder_asymptotics-2}) is applicable. In this Appendix we verify that contributions to these integrals from the regions where the modes have the asymptotic form (\ref{small-arg}) do not significantly modify the result that we obtain in (\ref{2-loop-current-leading}).

When, in the previous section, we substituted the asymptotics (\ref{cylinder_asymptotics-2}) into the integral (\ref{n_p(T)-2-loop-normal-g}), we obtained a sum of 256 terms over $ \alpha,\beta,\gamma,\delta,\alpha',\beta',\gamma',\delta'=\pm $ and identified the leading terms. However, if one considers narrow regions, where

\begin{equation}\label{small-zones}
|p_1+eE\tau_{1,2}|\lesssim m,\; |p_1-q_1+eE\tau_{1,2}|\lesssim m;\quad\quad
|k_1+eE\tau_{3,4}|\lesssim m,\; |k_1-q_1+eE\tau_{3,4}|\lesssim m,
\end{equation}
we have to use the asymptotics (\ref{small-arg}) for the corresponding mode functions, rather than (\ref{cylinder_asymptotics-2}).

For the leading two-loop integral (\ref{2-loop-n-p(T)-leading}), we found that it is proportional $ \delta(|\mathbf{q}|) $. This fact does not change if we replace some limits of the integration over $ \tau_{1,2,3,4} $ in accordance with  (\ref{small-zones}). The structure of $ C_\pm$-coefficients also does not change, from which it follows that non-oscillating contributions to the current are still cancel, as in (\ref{2-loop-current-leading-pre-final}). However, any replacement of the integration limits eliminates one power of $ T $. This leads to the fact that considering the regions (\ref{small-zones}) gives corrections of the order $ \mathcal{O}(T^3) $ or lower.

Then, we have to consider subleading terms in the sum of 256 terms, which we obtained after the substitution of the asymptotics (\ref{cylinder_asymptotics-2}) in (\ref{n_p(T)-2-loop-normal-g}). First, one can demonstrate, similarly to the derivation of the expression (\ref{2-loop-n-p(T)-leading}), that only such terms, that accumulate around $ |\mathbf{q}|=\mathcal{O}(1/T) $, survive. Further consideration is similar to Appendix (\ref{2-loop-Appendix}) and here is a brief summary:
\begin{itemize}
    \item $ \alpha'=\beta'=\gamma'=\delta' $ and $ \alpha'=\beta'=-\gamma'=-\delta' $: the integrand oscillates with $ X_{1,2}^2 $, where $ X_i=p_1+eE\tau_i $, thus the integrals over $ X_{1,2} $ do not grow with $ T $. The contribution to the current is of the order of $ \mathcal{O}\left(T^3\right) $ or lower.
    \item $ -\alpha'=\beta'=\gamma'=\delta',\; \alpha'=-\beta'=\gamma'=\delta',\; \alpha'=\beta'=-\gamma'=\delta' $, and $ \alpha'=\beta'=\gamma'=-\delta' $: the integrand oscillates only with $ X_1^2 $ (or only with $ X_2^2 $), thus only the integral over $ X_2 $ (or $ X_1 $) grows with $ T $. The integrand also oscillates over $ p_1 $ (this fact follows from the structure of $ C_\pm $-coefficients). Thus, the integral over $ p_1 $ does not grow with $ T $. In this case, the contribution to the current is also of the order of $ \mathcal{O}\left(T^3\right) $ or lower.
\end{itemize}
Similar results can be obtained for the one-loop correction to the Keldysh propagator of the photons. As a result, all subleading contributions to (\ref{2-loop-current-leading}) are of the order of $ \mathcal{O}\left(T^3\right) $ or lower.

\bibliography{bibliography}

\begin{thebibliography}{10}

\bibitem{Schwinger:1951nm}
Julian~S. Schwinger.
\newblock {On gauge invariance and vacuum polarization}.
\newblock {\em Phys. Rev.}, 82:664--679, 1951.

\bibitem{Grib:1976zw}
A.~A Grib, V.~M. Mostepanenko, and V.~M. Frolov.
\newblock {Particle Creation and Scattering by the Nonstationary
  Electromagnetic Field in Canonical Formalism}.
\newblock {\em Teor. Mat. Fiz.}, 26:221--233, 1976.

\bibitem{Grib:1980aih}
A.~A. Grib, S.~G. Mamayev, and V.~M. Mostepanenko.
\newblock {\em {Vacuum quantum effects in strong fields}}.
\newblock Friedmann Laboratory Publishing, St.Petersburg, 1994.

\bibitem{Zeldovich:1971mw}
Ya.~B. Zeldovich and Alexei~A. Starobinsky.
\newblock {Particle production and vacuum polarization in an anisotropic
  gravitational field}.
\newblock {\em Zh. Eksp. Teor. Fiz.}, 61:2161--2175, 1971.

\bibitem{Nikishov}
A.I. Nikishov.
\newblock {Problems of intense external-field intensity in quantum
  electrodynamics}.
\newblock {\em Journal of Soviet Laser Research}, 6:619–--717, 1985.

\bibitem{Nikishov:1969tt}
A.~I. Nikishov.
\newblock {Pair production by a constant external field}.
\newblock {\em Zh. Eksp. Teor. Fiz.}, 57:1210--1216, 1969.

\bibitem{Gavrilov:1996pz}
S.~P. Gavrilov and D.~M. Gitman.
\newblock {Vacuum instability in external fields}.
\newblock {\em Phys. Rev. D}, 53:7162--7175, 1996.

\bibitem{Kluger:1992gb}
Y.~Kluger, J.~M. Eisenberg, B.~Svetitsky, F.~Cooper, and E.~Mottola.
\newblock {Fermion pair production in a strong electric field}.
\newblock {\em Phys. Rev. D}, 45:4659--4671, 1992.

\bibitem{Gavrilov:2005dn}
S.~P. Gavrilov.
\newblock {Effective energy-momentum tensor of strong-field QED with unstable
  vacuum}.
\newblock {\em J. Phys. A}, 39:6407--6413, 2006.

\bibitem{Gavrilov:2007hq}
S.~P. Gavrilov and Dmitry~M. Gitman.
\newblock {One-loop energy-momentum tensor in QED with electric-like
  background}.
\newblock {\em Phys. Rev. D}, 78:045017, 2008.

\bibitem{Gavrilov:2007ij}
S.~P. Gavrilov and D.~M. Gitman.
\newblock {Energy-momentum tensor in thermal strong-field QED with unstable
  vacuum}.
\newblock {\em J. Phys. A}, 41:164046, 2008.

\bibitem{Gavrilov:2008fv}
S.~P. Gavrilov and D.~M. Gitman.
\newblock {Consistency restrictions on maximal electric field strength in QFT}.
\newblock {\em Phys. Rev. Lett.}, 101:130403, 2008.

\bibitem{Gavrilov:2012jk}
S.~P. Gavrilov, D.~M. Gitman, and N.~Yokomizo.
\newblock {Dirac fermions in strong electric field and quantum transport in
  graphene}.
\newblock {\em Phys. Rev. D}, 86:125022, 2012.

\bibitem{Anderson:2013zia}
Paul~R. Anderson and Emil Mottola.
\newblock {Quantum vacuum instability of
  \textquotedblleft{}eternal\textquotedblright{} de Sitter space}.
\newblock {\em Phys. Rev. D}, 89:104039, 2014.

\bibitem{Anderson:2013ila}
Paul~R. Anderson and Emil Mottola.
\newblock {Instability of global de Sitter space to particle creation}.
\newblock {\em Phys. Rev. D}, 89:104038, 2014.

\bibitem{Krotov:2010ma}
Dmitry Krotov and Alexander~M. Polyakov.
\newblock {Infrared Sensitivity of Unstable Vacua}.
\newblock {\em Nucl. Phys. B}, 849:410--432, 2011.

\bibitem{Akhmedov:2020dgc}
E.~T. Akhmedov, A.~V. Anokhin, and D.~I. Sadekov.
\newblock {Currents of created pairs in strong electric fields}.
\newblock {\em Int. J. Mod. Phys. A}, 36(19):2150134, 2021.

\bibitem{GribBook}
{A.A. Grib, S.G. Mamaev, and V.M. Mostepanenko}.
\newblock {\em {Quantum effects in intense strong fields}}.
\newblock Atomizdat, Moscow, 1980.

\bibitem{Akhmedov:2014hfa}
E.~T. Akhmedov, N.~Astrakhantsev, and F.~K. Popov.
\newblock {Secularly growing loop corrections in strong electric fields}.
\newblock {\em JHEP}, 09:071, 2014.

\bibitem{Akhmedov:2014doa}
E.~T. Akhmedov and F.~K. Popov.
\newblock {A few more comments on secularly growing loop corrections in strong
  electric fields}.
\newblock {\em JHEP}, 09:085, 2015.

\bibitem{Akhmedov:2023zfy}
E.~T. Akhmedov, P.~S. Zavgorodny, D.~I. Sadekov, and K.~A. Kazarnovskii.
\newblock {Loop corrections to the current of pairs created in a lengthy
  electric pulse}.
\newblock {\em Phys. Rev. D}, 107(12):125006, 2023.

\bibitem{Akhmedov:2024rkt}
E.~T. Akhmedov and P.~S. Zavgorodny.
\newblock {Higher loop corrections to the current of created pairs in the
  lengthy electric pulse}.
\newblock 9 2024.

\bibitem{Akhmedov:2021rhq}
E.~T. Akhmedov.
\newblock {Curved space equilibration versus flat space thermalization: A short
  review}.
\newblock {\em Mod. Phys. Lett. A}, 36(20):2130020, 2021.

\bibitem{kamenev2023field}
Alex Kamenev.
\newblock {\em Field theory of non-equilibrium systems}.
\newblock Cambridge University Press, 2023.

\bibitem{Akhmedov:2021vfs}
Emil~T. Akhmedov and Kirill Kazarnovskii.
\newblock {Thermalization with Non-Zero Initial Anomalous Quantum Averages}.
\newblock {\em Universe}, 8(3):162, 2022.

\bibitem{Akhmedov:2017ooy}
E.~T. Akhmedov, U.~Moschella, K.~E. Pavlenko, and F.~K. Popov.
\newblock {Infrared dynamics of massive scalars from the complementary series
  in de Sitter space}.
\newblock {\em Phys. Rev. D}, 96(2):025002, 2017.

\bibitem{Akhmedov:2024npw}
E.~T. Akhmedov, V.~I. Lapushkin, and D.~I. Sadekov.
\newblock {Light fields in various patches of de~Sitter spacetime}.
\newblock {\em Phys. Rev. D}, 111(12):125015, 2025.

\end{thebibliography}
\bibliographystyle{unsrt}
\end{document}